\documentclass[american,aps,pra,showpacs,preprint,tightenlines,superscriptaddress,showkeys]{revtex4}
\usepackage[T1]{fontenc}
\usepackage[latin9]{inputenc}
\usepackage{color}
\usepackage{longtable}
\usepackage{float}
\usepackage{amsmath}
\usepackage{amssymb}
\usepackage{graphicx}

\makeatletter

\providecommand{\tabularnewline}{\\}

\@ifundefined{textcolor}{}
{%
 \definecolor{BLACK}{gray}{0}
 \definecolor{WHITE}{gray}{1}
 \definecolor{RED}{rgb}{1,0,0}
 \definecolor{GREEN}{rgb}{0,1,0}
 \definecolor{BLUE}{rgb}{0,0,1}
 \definecolor{CYAN}{cmyk}{1,0,0,0}
 \definecolor{MAGENTA}{cmyk}{0,1,0,0}
 \definecolor{YELLOW}{cmyk}{0,0,1,0}
}

\usepackage{babel}

\@ifundefined{definecolor}{\@ifundefined{definecolor}
 {\@ifundefined{definecolor}
 {\usepackage{color}}{}
}{}
}{}

\@ifundefined{definecolor}{\@ifundefined{definecolor}
 {\@ifundefined{definecolor}
 {\@ifundefined{definecolor}
 {\usepackage{color}}{}
}{}
}{}
}{}
\usepackage{babel}

\@ifundefined{textcolor}{}{%
 \definecolor{BLACK}{gray}{0}
 \definecolor{WHITE}{gray}{1}
 \definecolor{RED}{rgb}{1,0,0}
 \definecolor{GREEN}{rgb}{0,1,0}
 \definecolor{BLUE}{rgb}{0,0,1}
 \definecolor{CYAN}{cmyk}{1,0,0,0}
 \definecolor{MAGENTA}{cmyk}{0,1,0,0}
 \definecolor{YELLOW}{cmyk}{0,0,1,0}
 }

\usepackage{epsfig}

\usepackage{amsfonts}

\usepackage{babel}

\addto\shorthandsspanish{\spanishdeactivate{~<>}}
\usepackage{babel}

\makeatother

\usepackage{babel}
\begin{document}

\title{Quantum state transfer in disordered spin chains: How much engineering
is reasonable?}

\author{Analia Zwick}

\affiliation{Department of Chemical Physics, Weizmann Institute of Science, 76100
Rehovot, Israel.}

\affiliation{Facultad de Matemática, Astronomía y Física and Instituto de F\'{i}sica
Enrique Gaviola, Universidad Nacional de Córdoba, 5000 Córdoba, Argentina.}

\affiliation{Fakultät Physik, Technische Universität Dortmund, D-44221 Dortmund,
Germany.}

\author{Gonzalo A. Álvarez}

\affiliation{Department of Chemical Physics, Weizmann Institute of Science, 76100
Rehovot, Israel.}

\affiliation{Fakultät Physik, Technische Universität Dortmund, D-44221 Dortmund,
Germany.}

\author{Joachim Stolze}

\affiliation{Fakultät Physik, Technische Universität Dortmund, D-44221 Dortmund,
Germany.}

\author{Omar Osenda}

\affiliation{Facultad de Matemática, Astronomía y Física and Instituto de F\'{i}sica
Enrique Gaviola, Universidad Nacional de Córdoba, 5000 Córdoba, Argentina.}

% \date{\today}


\begin{abstract}
{The transmission of quantum states through spin chains is an important
element in the implementation of quantum information technologies.
Speed and fidelity of transfer are the main objectives which have
to be achieved by the devices even in the presence of imperfections
which are unavoidable in any manufacturing process. To reach these
goals, several kinds of spin chains have been suggested, which differ
in the degree of fine-tuning, or engineering, of the system parameters.
In this work we present a systematic study of two important classes
of such chains. In one class only the spin couplings at the ends of
the chain have to be adjusted to a value different from the bulk coupling
constant, while in the other class every coupling has to have a specific
value. We demonstrate that configurations from the two different classes
may perform similarly when subjected to the same kind of disorder
in spite of the large difference in the engineering effort necessary
to prepare the system. We identify the system features responsible
for these similarities and we perform a detailed study of the transfer
fidelity as a function of chain length and disorder strength, yielding
empirical scaling laws for the fidelity which are similar for all
kinds of chain and all disorder models. These results are helpful
in identifying the optimal spin chain for a given quantum information
transfer task. In particular, they help in judging whether it is worthwhile
to engineer all couplings in the chain as compared to adjusting only
the boundary couplings.} 
 
\end{abstract}
\maketitle

\section{Introduction}

Quantum information processing \cite{NC01} relies on a number of
key elements of technology. The qubits (quantum bits) as elementary
units of quantum information are required to fulfill certain conditions
\cite{divincenzo_physical_2000}. The information contained in the
qubits or registers of qubits must be processed by quantum gates.
Since non-trivial quantum computers are likely to contain a large
number of quantum gates and registers, information must be transferred
between different elements of the computer. Quantum information is
encoded in states of single separate qubits or in entangled states
of several qubits. Transfer of single-qubit states over long distances
can be achieved by photons. For the short distances between different
elements of a quantum computer, however, photons do not provide the
most practical means of quantum state transfer. Instead, linear arrays
of suitably coupled qubits appear to be more promising. Using the
natural identification of a qubit with a spin-1/2 system, these arrays
have come to be known as {\em quantum spin chains}. Many different
spin chain models and quantum state transfer protocols have been studied
from different points of view during the last few years \cite{Bos07,NJ13}.
In the present study we want to focus on the state transfer in spin-1/2
XX chains with nearest-neighbor coupling in the presence of static
coupling disorder. While XX chains under suitable circumstances can
be used for the transfer of entangled multi-qubit states \cite{albanese_mirror_2004,karbach_spin_2005},
we restrict ourselves to single-qubit state transfer here.

Disorder can spoil quantum information transfer through spin chains
in many different ways. External fields fluctuating in time and/or
space may act on each spin separately, and the couplings connecting
the spins may deviate from their design values or even fluctuate in
time. In the present study all external fields are assumed to be strictly
zero and all couplings are assumed to be time-independent. Also, no
active control measures involving time-dependent fields or couplings
will be considered. The kind of disorder studied here is assumed to
arise from fabrication defects which are unavoidable in the process
of building an artificial quantum spin chain for quantum state transfer.
There are many coupling designs which yield perfect or near-perfect
quantum state transfer under ideal circumstances, and it is important
to assess the robustness of those designs against disorder.

The quality of transmission, that is, the similarity between the transmitted
state at one end of the spin channel and the received state at the
other end, is usually measured by the fidelity \cite{bose_quantum_2003}.
For a set of manufactured quantum spin chains, the best-case, worst-case,
or average fidelities may be relevant, depending on the situation
at hand. The situation we imagine is a future ``integrated quantum
circuit'' with a large number of quantum gates and a correspondingly
large number of spin chains connecting gates and registers. In that
case the average fidelity will be most relevant, provided there are
means to reroute the quantum information in order to circumvent ``weak
links'', {\em i.e.} low-fidelity spin chains in the network. This
point of view, focusing on the average fidelity, is common to many
existing studies \cite{chiara_perfect_2005,allcock_quantum_2009,petrosyan_state_2010,zwick_robustness_2011,zwick_spin_2012,oh_heisenberg_2011};
see, however, Ref. \cite{Nik13} for a different point of view.

A disordered chain is represented by a spin-chain Hamiltonian where
the exchange couplings become random variables that model the static
disorder affecting the interaction between spins \cite{chiara_perfect_2005,petrosyan_state_2010,ronke_effect_2011,zwick_robustness_2011,zwick_spin_2012,wang_FaultTolerance_2013}.
Solving the Hamiltonian eigenvalue problem for many realizations of
the disorder then leads, naturally, to an eigenvalue distribution
and probability distributions for the eigenvectors. Both probability
distributions, for the eigenvalues and eigenvectors, determine the
behavior of the averaged fidelity \cite{zwick_robustness_2011}, but
are up to now poorly understood. Moreover, apparently the same scaling
law for the decay of the averaged fidelity holds for a broad class
of spin couplings and two different models of static disorder \cite{zwick_spin_2012}.

Many protocols for high-fidelity quantum information transmission
have been developed \cite{bose_quantum_2003,albanese_mirror_2004,Plenio_high_2005,burgarth_efficient_2005,burgarth_optimal_2007,Gualdi_perfect_2008,allcock_quantum_2009,petrosyan_state_2010,Feldman2010,murphy_communication_2010,oh_heisenberg_2011,Venuti_Long-distance_2007,Venuti_Qubit_2007}.
In this work we focus on some of the main proposals based on the natural
Hamiltonian dynamics of spin chains with time-independent nearest-neighbor
couplings and without any external fields. These proposals can be
grouped into two classes depending on the extent to which the spin
couplings are being tuned, or {\em engineered}. The first class
comprises chains in which the state transmitted is exactly equal to
the state received \cite{christandl_perfect_2004,christandl_perfect_2005,karbach_spin_2005,kay_perfect_2006,kay_perfect_2010},
{\em i.e.} these systems provide {\em perfect state transfer}
(PST). PST can only be achieved if all couplings between spins are
properly adjusted, hence one might also classify these systems as
{\em fully engineered}. The second class of systems \cite{wojcik_unmodulated_2005,Feldman2010,banchi_optimal_2010,zwick_quantum_2011,yao_robust_2011,Banchi2011,banchi_long_2011,zwick_spin_2012,BFR+12,Linneweber2012,Lorenzo2013,Qin2013}
is characterized by a much smaller degree of engineering leading to
very good (but not perfect) state transfer. These systems require
only the adjustment of the boundary couplings of the chain, hence
they may be called {\em boundary-controlled} chains. Note, however,
that this term does not imply any active control of the system when
in use: the boundary couplings are fixed once and for all in the manufacturing
process. Since there is no unique prescription as to the ``right''
value of the boundary couplings, they may be optimized in different
ways to meet different objectives; for example, there is a trade-off
between fidelity and speed of transmission \cite{zwick_robustness_2011,zwick_spin_2012}.
In contrast to the PST systems we will call this class {\em optimizable
state transfer} (OST) systems %
\footnote{The word ``optimized'' in place of ``optimizable'' used earlier
\cite{zwick_spin_2012} suggests a unique optimum which actually does
not always exist.%
}. In the presence of static disorder in the exchange couplings, OST
systems were shown \cite{zwick_spin_2012,yao_robust_2011,Yao2013}
to offer reliable and robust state transmission. In particular we
showed \cite{zwick_spin_2012} that the quality of information transfer
in OST systems under perturbation by disorder can be comparable to
or even better than that of fully engineered PST systems. In order
to make the present study self-contained we review below some of that
earlier work, focusing on the most robust systems. We illustrate the
properties of the chains which ensure, on average, a successful state
transmission. From these properties, it can be understood why seemingly
different systems show qualitatively the same performance when subjected
to the same kind of disorder. In our numerical studies it turns out
that in all systems the transfer fidelity shows similar scaling laws
with respect to the length of the chain and the strength of the disorder.
Our results may help in identifying the optimal spin chain for a given
quantum information transfer task. In particular, they help in deciding
whether it is worthwhile to engineer all couplings in the chain as
compared to adjusting only the boundary couplings.

The paper is organized as follows. In Section~\ref{channels} we
present the Hamiltonian of the quantum spin chains whose performance
we want to compare, the transfer protocol and the disorder model.
In Section~\ref{eigen} we analyse the properties related to eigenvalues
and eigenvectors that result in a robust transfer of quantum states,
building in part on our earlier work \cite{zwick_robustness_2011,zwick_spin_2012}.
In Section~\ref{Chapter Fidelity} we study in detail how fast and
how reliably the different spin chains can transfer quantum information.
In a previous study \cite{chiara_perfect_2005} it was found that
the averaged fidelity $\overline{F}$ of a specific type of spin chain
of length $N$ depends on the scaling variable $N\varepsilon^{\beta}$,
where $\varepsilon$ is the strength of the disorder and $\beta=2$.
We find that other types of spin chains obey similar scaling laws.
In Section~\ref{Chapter Fidelity} we furthermore compare the different
systems and show how they can be grouped in several classes, before
we conclude with Section~\ref{conclusions}.

\section{Spin channels}

\label{channels} We consider two different types of spin chains for
state transfer: \emph{boundary-controlled} optimizable state-transfer
(OST) type \cite{wojcik_unmodulated_2005,zwick_quantum_2011,zwick_spin_2012,banchi_optimal_2010,banchi_long_2011},
and \emph{perfect state transfer} (PST) type \cite{albanese_mirror_2004,christandl_perfect_2004,karbach_spin_2005,christandl_perfect_2005,kay_perfect_2006,wang_all_2011,zwick_robustness_2011}.
Both are described by a XX Hamiltonian
\begin{equation}
H=\frac{1}{2}\sum_{i=1}^{N-1}J_{i}\left(\sigma_{i}^{x}\sigma_{i+1}^{x}+\sigma_{i}^{y}\sigma_{i+1}^{y}\right)\label{eq:hamiltonian XY gral}
\end{equation}
where $\sigma_{i}^{x,y}$ are the Pauli matrices, $N$ is the chain
length, and $J_{i}$ are the time-independent exchange interaction
couplings between neighboring spins. The $J_{i}$ are allowed to vary
in space, but we assume mirror symmetry with respect to the center
of the chain, $J_{i}=J_{N-i}$.

The\emph{ }boundary-controlled spin chains are a mono-parametric family
of chains, such that
\begin{equation}
J_{1}=J_{N-1}=\alpha J,\quad\mbox{and}\quad J_{i}=J,\;\forall i\neq1,N-1.\label{eq:Hamilt alpha}
\end{equation}
The parameter $\alpha$ modifies the strength of the exchange interaction
of the boundary spins $i\!=\!\!1$ and $i\!=\!\! N$ with their respective
nearest-neighbor spins, otherwise the chains are homogeneous.

In contrast, PST spin chains are designed to allow for perfect state
transmission at some time. That calls for a certain structure of the
energy spectrum, and by solving an inverse eigenvalue problem the
corresponding values of the couplings $J_{i}$ can be determined.
We refer to these chains as {\em fully engineered} since all $N\!-\!1$
couplings $J_{i}$ must be adjusted properly.

\subsection{Protocol and fidelity of state transmission}

The goal is to transmit a quantum state $\left|\psi_{0}\right\rangle $
initially stored on the first spin ($i=1$) to the last spin of the
chain ($i=N$). $\left|\psi_{0}\right\rangle $ is an arbitrary normalized
superposition of the spin down ($\left|0\right\rangle $) and up ($\left|1\right\rangle $)
states of the first spin, with the remaining spins of the chain initialized
in a spin down state %
\footnote{In fact, only the spin $i=1$ has to be initialized, while the other
spins can be in arbitrary states in the beginning. This is because
the Hamiltonian of Eq. (\ref{eq:hamiltonian XY gral}) can be mapped
into a non-interacting fermion Hamiltonian with nearest neighbor hopping
with the Jordan-Wigner transformation \cite{Lieb1961}. We have assumed
the spins $i>1$ to be initialized exclusively for the ease of discussion. %
}. The Hamiltonian (\ref{eq:hamiltonian XY gral}) conserves the number
of up spins because $[H,\Sigma_{i}\sigma_{i}^{z}]=0$. Therefore the
component of the initial state $|\mathbf{0}\rangle=|00...0\rangle$
is an eigenstate of $H$ and only the component $\mathbf{|1}\rangle=\mathbf{|}1_{1}0......0\rangle$
evolves within the one excitation subspace spanned by the basis states
$\mathbf{|i}\rangle=\mathbf{|}0...01_{i}0...0\rangle$. To evaluate
how well an unknown initial state is transmitted, we use the transmission
fidelity, averaged over all possible $\left|\psi_{0}\right\rangle $
from the Bloch sphere
\begin{eqnarray}
F(t) & = & \frac{|f_{N}(t)|}{3}\cos\gamma+\frac{|f_{N}(t)|^{2}}{6}+\frac{1}{2},\label{eq:Averaged-Fidelity}
\end{eqnarray}
where $|f_{N}(t)|^{2}=\left|\langle\mathbf{N}|e^{-\frac{iHt}{\hslash}}|\mathbf{1}\rangle\right|^{2}$
is the fidelity of transfer between states $|\mathbf{1}\rangle$ and
$|\mathbf{N}\rangle$ and $\gamma=\arg\left|f_{N}(t)\right|$ \cite{bose_quantum_2003}.
Because the phase $\gamma$ can be controlled by an external field
once the state is transferred, we consider $\cos\gamma=1$. By the
symmetries of the system, this fidelity can be expressed in terms
of the single-excitation energies $E_{k}$ and the eigenvectors $|\Psi_{k}\rangle$
of $H$, in the following way
\begin{equation}
|f_{N}(t)|^{2}=\sum_{k,s}(-1)^{k+s}P_{k,1}P_{s,1}e^{-i(E_{k}-E_{s})t/\hbar}\label{eq:fn}
\end{equation}
where $P_{k,1}=|\langle{\Psi_{k}}|{\mathbf{1}}\rangle|^{2}$ are the
eigenvector occupation probabilities on the first site of the chain.

\subsection{Static disorder models}

Static disorder in the couplings within the transfer channel is described
by $J_{i}\rightarrow J_{i}+\Delta J_{i}$ $(i=2,...,N-2)$ with $\Delta J_{i}$
a random variable. We consider two possible coupling disorder models:\textbf{
(a)} \textit{relative static disorder,} where each coupling is allowed
to fluctuate by a certain fraction of its ideal size, $\Delta J_{i}=J_{i}\delta_{i}$
\cite{chiara_perfect_2005,petrosyan_state_2010,zwick_robustness_2011,zwick_spin_2012},
and \textbf{(b)} \textit{absolute static disorder,} where all couplings
may fluctuate within a certain fixed range which we measure in terms
of $J_{max}=\max J_{i}$: $\Delta J_{i}=J_{max}\delta_{i}$ \cite{ronke_effect_2011,zwick_spin_2012}.
Each $\delta_{i}$ is an independent and uniformly distributed random
variable in the interval $\left[-\varepsilon_{J},\varepsilon_{J}\right]$.
$\varepsilon_{J}>0$ characterizes the strength of the disorder. The
two coupling disorder models are equivalent for the \emph{boundary-controlled}
spin chains since all couplings are equal there for $i=2,...,N-2$.
However, in the fully engineered PST systems $J_{max}-J_{min}$ depends
on the type of system and tends to increase with $N$ so that {absolute
disorder} is expected to be more damaging than relative disorder
in these systems. Which kind of disorder is relevant depends on the
particular experimental method used to engineer the spin chains \cite{ladd_quantum_2010}.

\section{Boundary controlled and fully engineered channels for robust state
transfer}

\label{antecedents}The boundary-controlled chains with optimizable
state transfer (OST), Eqs. (\ref{eq:hamiltonian XY gral}-\ref{eq:Hamilt alpha}),
can be optimized in two different ways \cite{zwick_spin_2012}. If
transmission speed is not an issue while high fidelity is desired,
a weak coupling $\alpha J=\alpha_{0}J\ll\frac{1}{\sqrt{N}}$ should
be chosen \cite{wojcik_unmodulated_2005,zwick_quantum_2011,zwick_spin_2012,yao_robust_2011}.
Perfect state transmission can then be obtained asymptotically for
vanishing $\alpha$ but the transfer time increases with decreasing
$\alpha$ and depends strongly on the parity of $N$, the spin chain
length \cite{wojcik_unmodulated_2005}. If speed is critical while
fidelity need not be perfect, the boundary coupling should be optimized
depending on the spin chain length, choosing $\alpha J=\alpha_{opt}J\backsimeq1.05N^{-\frac{1}{6}}$
\cite{banchi_optimal_2010,zwick_quantum_2011,zwick_spin_2012}. We
will denote these two cases by the symbols $\alpha_{0}$ and $\alpha_{opt}$,
respectively. 

\begin{figure}
\centering{}\includegraphics[scale=0.41]{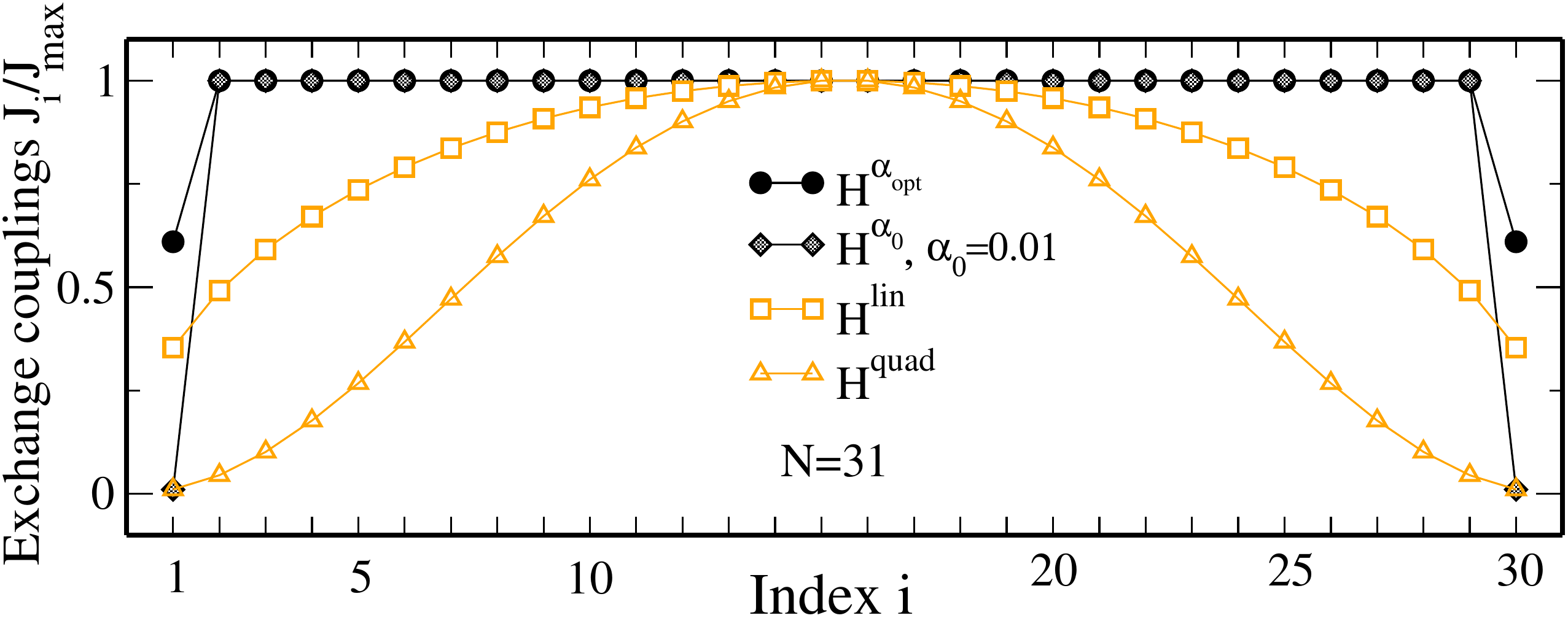} \caption{\label{fig:Ji-N31}(Color online) Exchange couplings $\frac{J_{i}}{J_{max}}$
for the boundary-controlled OST chains (black symbols) and for the
fully-engineered \emph{linear} and \emph{quadratic} PST chains (open
orange symbols). The couplings of the linear PST chain are known \cite{christandl_perfect_2004}
to show a circular pattern, those for the quadratic channel are determined
by solving an inverse eigenvalue problem. Chain length is $N=31$.}
\end{figure}

The class of fully engineered spin chains with perfect state transfer
(PST) is large; in fact there are countless ways to design PST-type
chains \cite{karbach_spin_2005,wang_all_2011,kay_perfect_2006}.
We analyzed the performance of PST chains with power-law energy spectrum
\cite{zwick_robustness_2011,zwick_spin_2012} under the influence
of static disorder. It turned out that systems with a \emph{linear}
or \emph{quadratic} energy spectrum (see Sec. \ref{eigen} below for
details) were most robust against static disorder. The fidelity (\ref{eq:fn})
of the state transfer and its robustness depend crucially on the probability
$P_{k,1}=|\langle\mathbf{1}|\Psi_{k}\rangle|^{2}$ and the shape of
some regions of the energy spectrum \cite{zwick_robustness_2011,zwick_spin_2012}.

In the present work we study the performance of two PST chains, denoted
by $^{lin}$ and $^{quad}$, respectively, in comparison to the OST
systems $^{\alpha_{opt}}$ and $^{\alpha_{0}}$ introduced above.
In the quadratic PST and weak-coupling OST cases the parity of $N$
will turn out to be important \cite{my64}. This similarity, along
with other similarities between OST and PST chains, can be understood
from properties of the eigenstates and eigenvalues, but by no means
directly from the distribution of the couplings $J_{i}$ which is
shown in Figure \ref{fig:Ji-N31}.

\subsection{Spectral properties\label{eigen}}

The key to high state transfer fidelity, as obvious from Eq.~(\ref{eq:fn}),
is in the spectrum of energies $E_{k}$ and the probabilities $P_{k,1}$.
These quantities help to understand \emph{(i)} similarities between
ideal chains from the PST and OST classes, respectively, and \emph{(ii)
}the differences in robustness against static disorder. Note that
\emph{(i)} is significant in terms of experimental feasibility, since
OST-chains are potentially easier to manufacture than PST-chains.

The single-excitation energy eigenvalues of the systems considered
are given as follows:

\emph{(a)} for OST-chains, $E_{k}=2J\,\cos\,\gamma_{k}$, where $k$
is given by the $N$ solutions of $\pm\cot\,\gamma_{k}\:\cot^{\pm1}(\frac{N-1}{2}\gamma_{k})=\frac{\alpha^{2}J{}^{2}}{2-\alpha^{2}J{}^{2}}$
\cite{wojcik_unmodulated_2005}; and

\emph{(b)} for PST-chains, $E_{\tilde{k}}\!=\!\frac{\pi\hbar}{\tau_{pst}}\text{sgn}(\tilde{k})|\tilde{k}|^{m},$
where $\tilde{k}\!=\!-\frac{{N-1}}{2},...,\frac{{N-1}}{2}$ and the
exponent $m\!=\!1$ for the \emph{linear} case and $m\!=\!2$ for
the \emph{quadratic} one, for odd $N$. If $N$ is even, $\tilde{k}\!=\!-\frac{N}{2},...,\frac{N}{2}$,
excluding zero, and in the linear (quadratic) case, $|\tilde{k}|$
($|\tilde{k}|^{2}$) has to be replaced by$|\tilde{k}|-\frac{1}{2}$
$\left(|\tilde{k}^{2}-\frac{1}{2}|\right)$ in $E_{\tilde{k}}$. Here,
$\tau_{pst}$ is the time after which the first perfect state transfer
occurs.

\begin{figure}
\centering{}\includegraphics[scale=0.41]{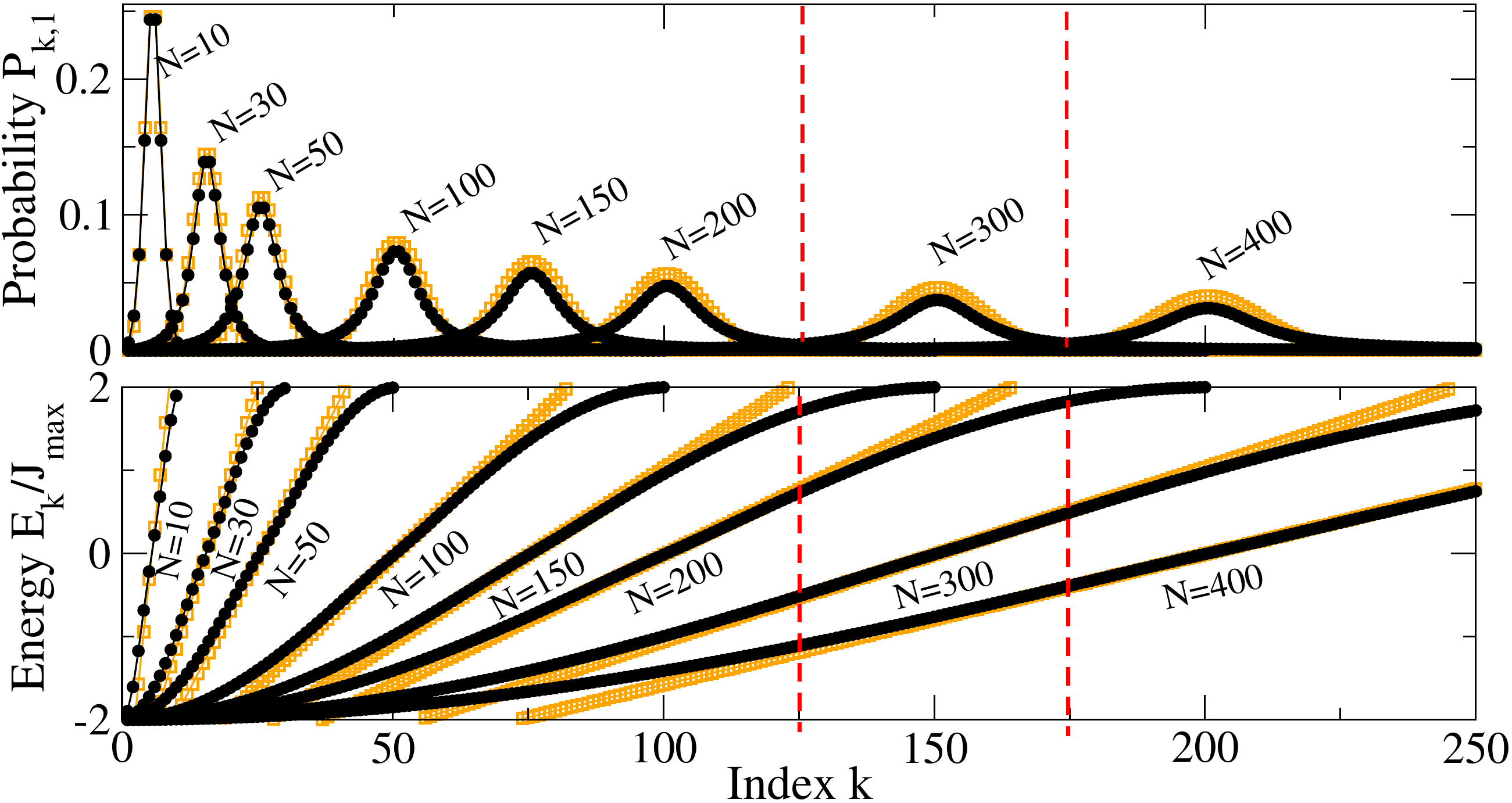} \caption{\label{fig:Pk1-Ek-N-Lin-opt} (Color online) Probabilities $P_{k,1}$
to find the initial state $\left|\psi_{0}\right\rangle =|\mathbf{1}\rangle$
in the eigenstate $\left|\Psi_{k}\right\rangle $, and energies $E_{k}$
of the systems $H^{\alpha_{opt}}$ (black solid dots) and $H^{lin}$
(orange open squares) for different chain lengths $N$. The dashed
vertical lines show, as an example for $N=300$, the region of dominant
energy eigenstates $|\Psi_{k}\rangle$ that contribute to the state
transfer. The energy spectrum in the relevant region is linear for
both systems. To stress the similarity between the energy spectra
the $E_{k}$ of $H^{lin}$ have been multiplied by $\frac{\pi}{2}$.}
\end{figure}

\begin{figure}
\centering{}\includegraphics[scale=0.41]{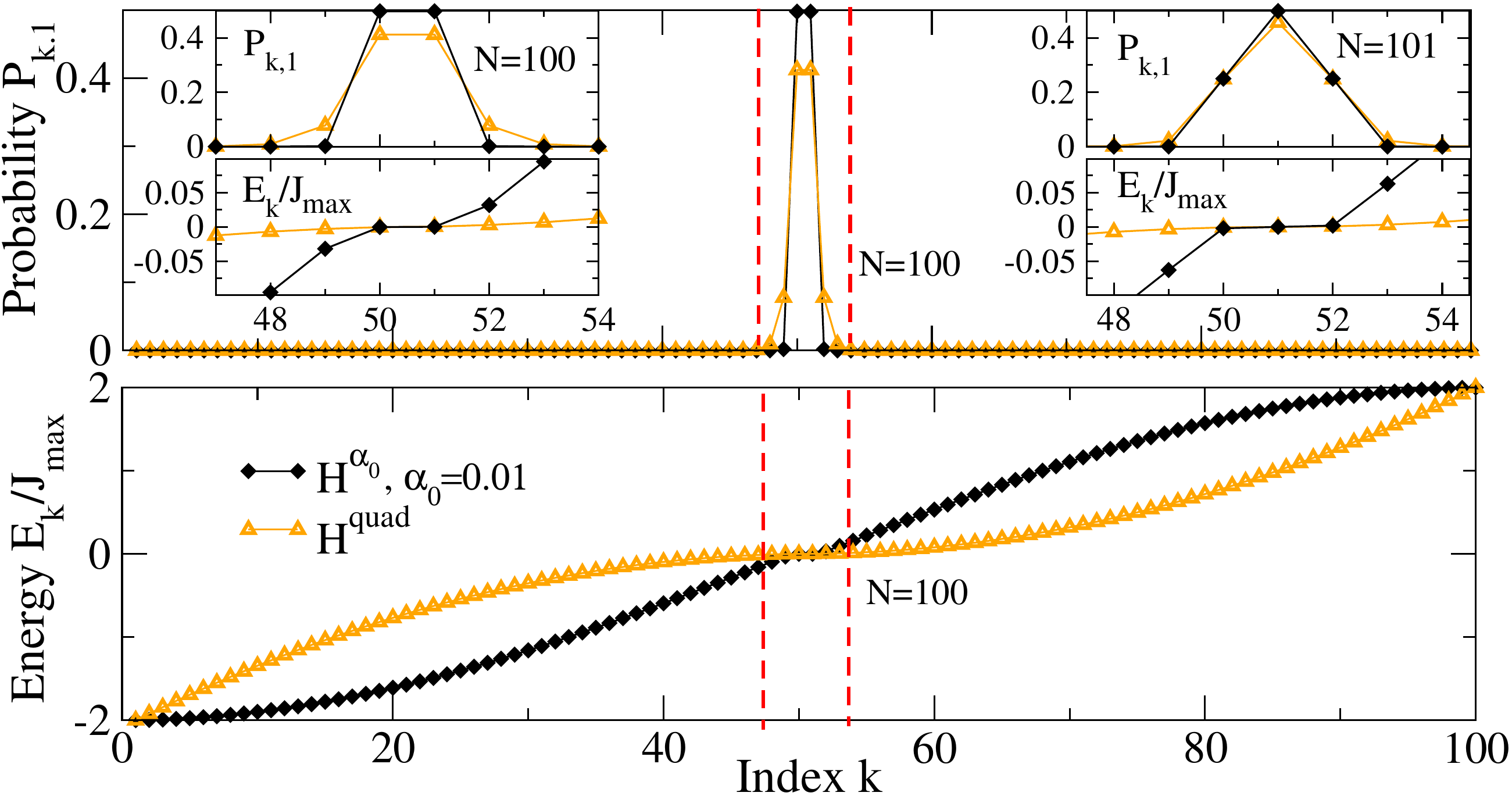} \caption{\label{fig:Pk1-Ek-Quad-alp0} (Color online) Probabilities $P_{k,1}$
to find the initial state $\left|\psi_{0}\right\rangle =|\mathbf{1}\rangle$
in the eigenstate $|\Psi_{k}\rangle$ and energies $E_{k}$ of the
system $H^{\alpha_{0}}$ ($\alpha=0.01$; black solid diamonds) and
$H^{quad}$ (orange open triangles) for even $N=100$. The dashed
vertical lines show the small region of dominant energy eigenstates
$|\Psi_{k}\rangle$ that contribute to the state transfer. This region
is shown magnified in the left inset. The right inset shows the same
for odd $N=101$.}
\end{figure}

A common property of all these systems that makes them robust is that
the eigenstates involved in the state transmission belong to the center
of the energy band \cite{zwick_robustness_2011,zwick_spin_2012}.
The $\alpha_{opt}$-OST chain has a linear spectrum in this energy
region as shown in Fig. \ref{fig:Pk1-Ek-N-Lin-opt} for several chain
lengths $N$. In contrast, the $\alpha_{0}$-OST chain has a rather
flat spectrum there, similar to the \emph{quadratic}-PST channel as
shown in Fig. \ref{fig:Pk1-Ek-Quad-alp0}.

At the same time, the probability of the $k$-th energy eigenstate
to participate in the state transfer is given by $P_{k,1}$, Eq. (\ref{eq:fn}).
For

\emph{(a)} OST-chains, $P_{k,1}^{\alpha_{opt}}$ is a Lorentzian distribution
\cite{banchi_long_2011} while $P_{k,1}^{\alpha_{0}}$ is essentially
non-zero only for two (three) values of $k$ when $N$ is even (odd)
\cite{wojcik_unmodulated_2005} (Fig. \ref{fig:Pk1-Ek-Quad-alp0});
whereas for

\emph{(b)} PST-chains, $P_{k,1}^{lin}$ follows a Gaussian distribution
(see Appendix \ref{sec: Appendix B. Gaussian Pk1}) while $P_{k,1}^{quad}$
is significantly different from zero only for two (three) values of
$k$ when $N$ is even (odd) (Fig. \ref{fig:Pk1-Ek-Quad-alp0}). The
significant contributions of $P_{k,1}$ for all the above channels
are thus all concentrated near the center of the energy band. The
varying degree of that concentration naturally separates the four
systems into two pairs: The $\alpha_{opt}$ and linear systems show
a broad $P_{k,1}$ distribution (Fig. \ref{fig:Pk1-Ek-N-Lin-opt}),
whereas in the $\alpha_{0}$ and quadratic systems, $P_{k,1}$ is
very narrow. Given the similarities in the relevant parts (as determined
by $P_{k,1}$ being of appreciable size) of the energy spectra $E_{k}$,
it is clear that the members of each pair can be expected to show
very similar behavior.

Figure \ref{fig:Pk1-Ek-N-Lin-opt} shows clearly the mechanism behind
the high state transfer fidelities shown by the two systems represented
there. While it takes a relatively large number of energy eigenstates
$|\Psi_{k}\rangle$ to resolve the localized initial state $|\psi_{0}\rangle$,
the eigenvalues $E_{k}$ of those states are equidistant due to the
linear spectrum. That makes the time evolution of the initial state
periodic with a period equal to twice the time needed for the state
transfer from site 1 to site $N$. Spatial symmetry then is the second
ingredient needed to provide for perfect or near-perfect state transfer.
The probability distributions $P_{k,1}$ in Fig. \ref{fig:Pk1-Ek-N-Lin-opt}
are known. $P_{k,1}^{\alpha_{opt}}$ was shown to be Lorentzian \cite{banchi_long_2011}:
\begin{equation}
P_{k,1}^{\alpha_{opt}}\simeq\frac{1}{\pi}\frac{\Gamma}{(k-k_{0})^{2}+\Gamma^{2}},
\end{equation}
where $k_{0}=\frac{N+1}{2}$ and $\Gamma\simeq(\frac{10}{N})^{-0.63}$
(best fit for $N=400$). $P_{k,1}^{linear}$ is asymptotically Gaussian
(see Appendix \ref{sec: Appendix B. Gaussian Pk1})
\begin{equation}
P_{k,1}^{linear}\simeq Ae^{-\frac{(k-k_{0})^{2}}{2\sigma^{2}}},
\end{equation}
with $A=\frac{0.8}{\sqrt{N}}$ and $\sigma=\frac{2}{\sqrt{N}}$ (for
$N=400$).

The $\alpha_{0}$-OST and\emph{ quadratic}-PST systems from Fig. \ref{fig:Pk1-Ek-Quad-alp0}
are quite different. There the state transfer is performed by a very
small (and $N$-independent) number of eigenstates. For even $N$,
in the limit $\alpha_{0}\rightarrow0^{+}$ the dominant eigenvectors
belong to the two energies $E_{k_{\pm}}=\pm|E|_{min}$ closest to
zero, that is, $|\Psi_{k_{\pm}}\rangle$ with $k_{-}=\frac{N}{2},\: k_{+}=\frac{N}{2}+1$.
The probabilities for these states are $P_{k_{\pm},1}^{\alpha_{0}}\simeq\frac{1}{2}$,
and, by normalization $P_{k,1}^{\alpha_{0}}\simeq0$ for $k\neq k_{\pm}$.
For odd $N$, in contrast, three eigenvectors, $|\Psi_{k_{\pm}}\rangle$
and $|\Psi_{k_{0}}\rangle$ (with $E_{k_{0}}^{\alpha_{0}}=0$) are
dominant, with $P_{k_{\pm},1}^{\alpha_{0}}\simeq\frac{1}{4}$ and
$P_{k_{0},1}^{\alpha_{0}}\simeq\frac{1}{2}$. The nonzero energy eigenvalues
for small $\alpha_{0}$ are very different: $E_{k_{\pm},even}^{\alpha_{0}}\sim\pm\alpha_{0}^{2}$
and $E_{k_{\pm},odd}^{\alpha_{0}}\sim\pm\frac{2\alpha_{0}}{\sqrt{N}}$
\cite{wojcik_unmodulated_2005}. For \emph{quadratic}-PST systems
we numerically observe a similar behavior with the addition of a small
contribution from two more eigenstates, those belonging to the pair
of energies next closest to zero.

We have analyzed the changes in both the energy spectrum $E_{k}$
and the structure of the eigenstates, as displayed by the probabilities
$P_{k,1}$, under the influence of disorder \cite{zwick_robustness_2011,my64}.
As a general rule it turns out that energies near the center of the
energy band are least affected by disorder in the class of spin chains
discussed here. Since the band-center states are most important for
state transfer, this sounds like a piece of good news. Of all systems,
$H^{quad}$ shows the smallest ``spectral sensitivity'', as measured
by the standard deviation of $E_{k}$ for given disorder strength
$\varepsilon_{J}$. However, since both the energy eigenvalues and
the occupation probabilities $P_{k,1}$ influence the fidelity, this
does not mean that $H^{quad}$ is the most robust state transfer system
under all circumstances. Quite to the contrary, $H^{quad}$ tends
to be rather robust for odd $N$ and quite delicate for even $N$;
see Section \ref{disordered_fidelity} for details.

Below we shall discuss the performance of all channels as measured
by the transfer time and the transfer fidelity (\ref{eq:Averaged-Fidelity})
and we shall observe marked similarities between the members of each
of the two pairs of state transfer channels. These similarities can
be explained by the features of the eigenvalues $E_{k}$ and probabilities
$P_{k,1}$ just discussed.

\section{Performance comparison: Transfer time and fidelity\label{Chapter Fidelity}}

\subsection{The transfer time\label{sub:The-transfer-time}}

\begin{figure}
\begin{centering}
\includegraphics[scale=0.25]{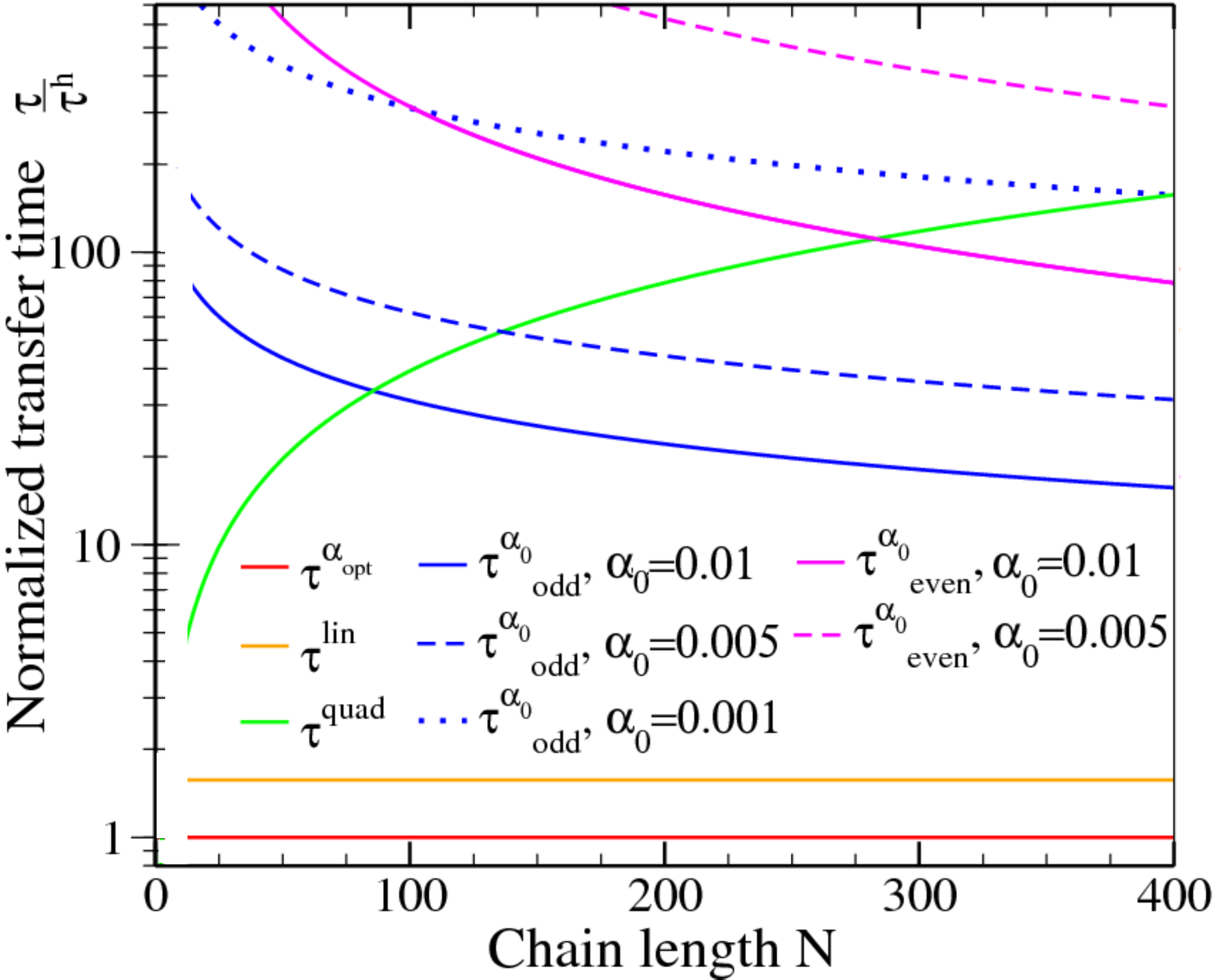} 
\par\end{centering}

\caption{\label{transfer-time-N-alp} (Color online) Transfer times $\tau$
where the maximum transfer fidelity is achieved as a function of $N$
and $\alpha_{0}$. $\tau^{h}$ is the transfer time for a homogeneous
chain, \textit{ie}. $J_{i}=J$ $\forall J_{i}$. Values $\frac{\tau}{\tau^{h}}(N,\alpha_{0})$
are given for $\alpha_{0}=0.001,0.005,0.01.$ }
\end{figure}

There is no unique way to define the transfer time for arbitrary state
transfer channels, since the fidelity as a function of time may show
a complicated pattern of maxima \citep{zwick_robustness_2011}. As
a working definition we may say that the transfer time is defined
by the first maximum of useful size in the fidelity. In the examples
discussed here the fidelity does not show erratic dynamics and the
meaning of the transfer time will be unambiguous. The transfer time
$\tau$ depends strongly on the type of spin chain \citep{wojcik_unmodulated_2005,yung_quantum_2006,murphy_communication_2010,zwick_robustness_2011,zwick_quantum_2011}.
PST channels have commensurate energies $E_{k}$; that means, all
transition frequencies share a common divisor $\tau_{PST}$ to make
$f_{N}\!=\!1$ in Eq. (\ref{eq:fn}) \citep{christandl_perfect_2004,karbach_spin_2005}.
In particular,\emph{ }the times\emph{ $\tau^{lin}$} and \emph{$\tau^{quad}$}
for \emph{linear} and \emph{quadratic} PST channels are half of the
mesoscopic echo time \citep{zwick_robustness_2011}, which is the
characteristic (round-trip) time of the information propagation within
the chain. For other systems the transfer time is generally longer
\citep{zwick_robustness_2011}. \textcolor{black}{For some PST chains
the exact transfer time can be obtained analytically \citep{karbach_spin_2005,albanese_mirror_2004,christandl_perfect_2004,kay_perfect_2010,zwick_robustness_2011}.
For other types of chains, such as the boundary-controlled ones considered
in this work, the transfer time must be obtained by }\textcolor{black}{\emph{ad
hoc}}\textcolor{black}{{} means \citep{wojcik_unmodulated_2005,banchi_long_2011,zwick_quantum_2011}.}
\begin{table}
\begin{longtable}{|c|c|c|c|c|c|}
\hline 
Homogeneous  & \multicolumn{3}{c|}{$\alpha$-OST channel} & \multicolumn{2}{c|}{PST channel}\tabularnewline
\hline 
$\tau^{h}\simeq\frac{N}{2J_{max}}$  & $\tau^{\alpha_{opt}}\gtrsim\frac{N}{2J_{max}}$  & $\tau_{odd}^{\alpha_{0}}\simeq\frac{\pi\sqrt{N-2}}{2\alpha J_{max}}$  & $\tau_{even}^{\alpha_{0}}\simeq\frac{\pi}{2\alpha^{2}J_{max}}$  & $\tau^{lin}\simeq\frac{\pi N}{4J_{max}}$  & $\tau_{odd,even}^{quad}\simeq\frac{\pi N^{2}}{8J_{max}}$\tabularnewline
\hline 
$\simeq\tau^{h}$  & $\gtrsim\tau^{h}$  & $\simeq\frac{\pi\sqrt{N-2}}{\alpha N}\tau^{h}$  & $\simeq\frac{\pi}{\alpha^{2}N}\tau^{h}$  & $\simeq\frac{\pi}{2}\tau^{h}$  & $\simeq\frac{\pi N}{8}\tau^{h}$\tabularnewline
\hline 
\end{longtable}\caption{\label{Table_transfer-time2} Transfer times $\tau$ where the maximum
of the fidelity of transmission is obtained for different spin chains.
The last row compares these times with the transfer time $\tau^{h}$
of a homogeneous chain, \textit{ie}. $J_{i}=J$ $\forall J_{i}$.
The transfer time $\tau^{lin}$ was obtained in Ref. \cite{christandl_perfect_2004},
$\tau^{\alpha_{0}}$ in Ref. \cite{wojcik_unmodulated_2005}, $\tau^{\alpha_{opt}}$
in Ref. \cite{zwick_quantum_2011} and we obtained $\tau_{odd,even}^{quad}$
from our numerical results.}
\end{table}

The transfer times for the chains considered here are listed in Table
\ref{Table_transfer-time2}. The shortest transfer time is achieved
by the boundary-controlled chain working in the \emph{optimal} regime
\citep{zwick_spin_2012}. This transfer time is very close to the
bound given by the quantum speed limit $\tau^{h}=\frac{N}{2J_{max}}$
given by the maximum group velocity of excitations in the homogeneous
chain \citep{lieb_finite_1972,yung_quantum_2006,levitin_fundamental_2009,murphy_communication_2010}.
The transfer times for different channels are given in units of $\tau^{h}$
in the last row of Table \ref{Table_transfer-time2}. The shortest
time $\tau^{\alpha_{opt}}$ is followed by $\tau^{lin}$, and then
by the remaining transfer times, $\tau^{\alpha_{0}}$ and $\tau^{quad}$
which depend on $N$ and $\alpha_{0}$ as shown in Fig. \ref{transfer-time-N-alp}.
If $\alpha_{0}\ll\frac{1}{\sqrt{N}}$, we always have $\tau_{odd}^{\alpha_{0}}<\tau_{even}^{\alpha_{0}}$
and for $\alpha\lesssim\frac{8\sqrt{N-2}}{N^{3/2}}$, $\tau^{quad}<\tau_{odd}^{\alpha_{0}}$.

\begin{figure}
\centering{}\includegraphics[scale=0.25]{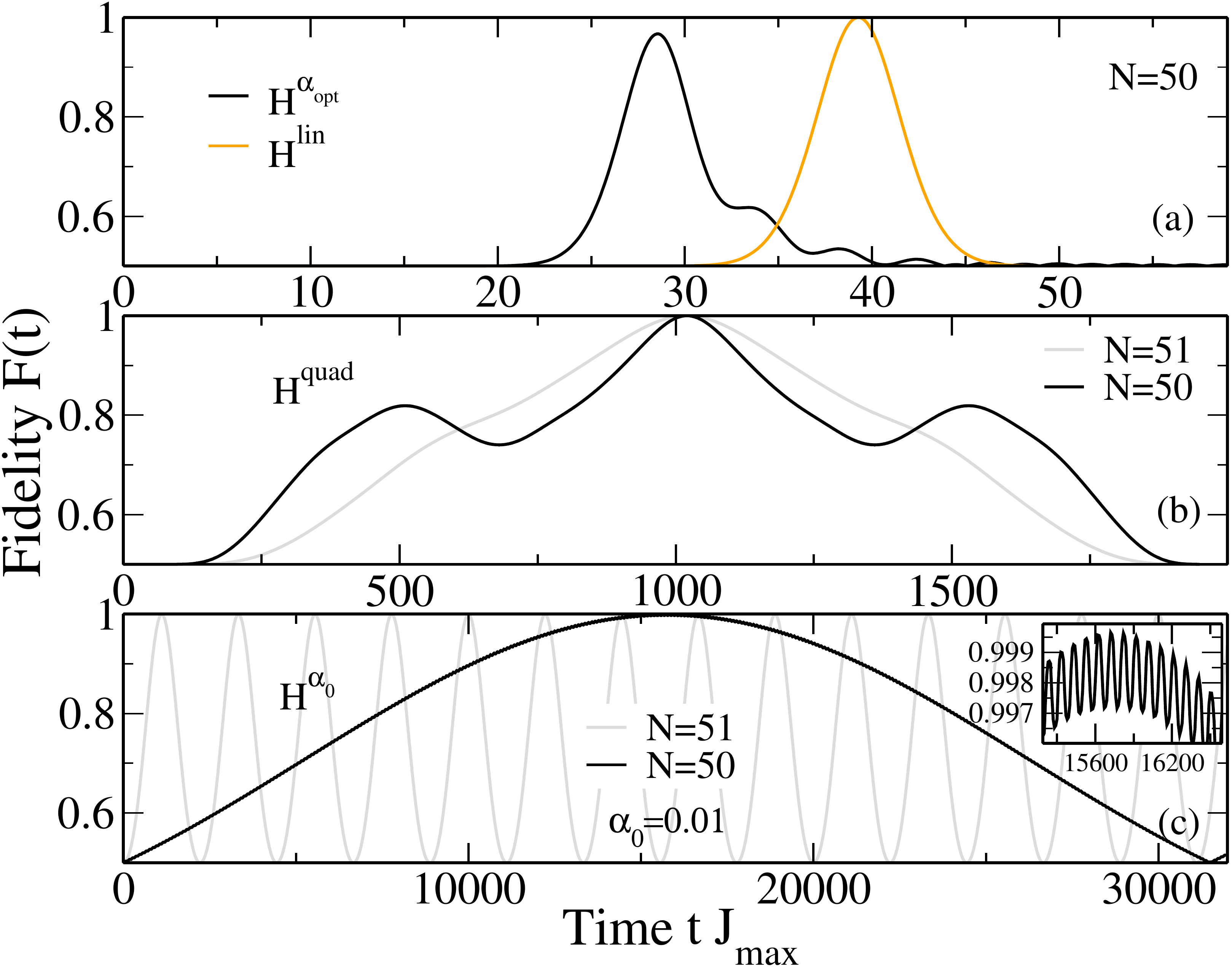} \caption{\label{fig:F-vs-t-Quadratic-alphachico-Npar-impar}(Color online)
Transmission fidelity $F(t)$ as a function of time $t$. \textbf{(a)}
$F{}^{\alpha_{opt}}(t)$ (black line) and $F{}^{lin}(t)$ (orange
line) for an even chain length $N=50$. The behavior of the fidelity
is not affected by the parity of $N$. However, for \textbf{(b)} $F{}^{quad}(t)$
and \textbf{(c)} $F{}^{\alpha_{0}}(t)$, the parity of $N$ matters.
Panels (b,c) show the fidelity for $N=50$ (black line) and $N=51$
(gray line). The inset in (c) shows the oscillation observed for even
$N$ that is an evidence of the off-resonance transmission which is
much slower than the on-resonance transmission for odd $N$.}
\end{figure}

A comparison of the transfer fidelity as a function of time in the
absence of disorder is displayed in Fig. \ref{fig:F-vs-t-Quadratic-alphachico-Npar-impar}.
We can observe there the basic characteristics of each channel. In
particular, in Fig. \ref{fig:F-vs-t-Quadratic-alphachico-Npar-impar}a
we observe the faster transfer of the $\alpha_{opt}$-OST channel
as compared to the \emph{linear} one, while its fidelity maximum is
lower than that of the \emph{linear} PST-channel. For the $\alpha_{0}$-OST
and the \emph{quadratic} PST channels, the transfer is slower than
in the previous cases and it depends on the parity of the chain length.
This is because for odd $N$ the transmission of the state from the
boundary spins is mainly performed through an eigenstate of the bulk
spins $(i\!=\!2,..,N\!-\!1)$ that is in the center of the band ($E_{k}=0$)
and consequently on resonance with the boundary spins \citep{wojcik_unmodulated_2005,yao_robust_2011}.
However, for even $N$, the transfer proceeds through two eigenstates
of the bulk spins with finite energy and consequently off-resonance
with the boundary spins \citep{wojcik_unmodulated_2005,yao_robust_2011}.
The fast oscillation observed in the inset of Fig. \ref{fig:F-vs-t-Quadratic-alphachico-Npar-impar}c
for $F^{\alpha_{0}}$ with even $N$ is an evidence of the off-resonance
transmission. A simple calculation considering the dynamics in the
space spanned by the dominating eigenstates reveals the frequency
of the oscillation that varies with $N$ as $\omega_{\alpha_{0}}\sim\frac{\pi J_{max}}{N}\sqrt{\alpha_{0}^{2}N+1}$
and the amplitude as $A_{\alpha_{0}}\sim\frac{1}{4}\alpha_{0}^{2}N(N\alpha_{0}^{2}+1)^{-1}$.
The main slower oscillation that produces the state transfer comes
from only two eigenvectors of the total system $(i=1,..,N)$ strongly
localized in the channel's ends \citep{wojcik_unmodulated_2005}.
For odd $N$ the transmission is smooth because it proceeds through
the eigenstate with zero energy. A similar behaviour with respect
to the parity of $N$ is also observed in the \emph{quadratic}-PST
channel.

A second important time scale apart from the transfer time is what
one may call the window-time, that is, the width of the fidelity maximum
which defines the transfer time. That time scale defines the precision
in time that is needed to read out the state transferred with high
fidelity. While the \emph{quadratic} and $\alpha_{0}$ channels are
slower in transfer than their \emph{linear} and $\alpha_{opt}$ counterparts,
they have the widest window of time. The achievable transfer fidelity
at time $\tau$ will be discussed below, when we deal with state transfer
in the presence of imperfections.

\subsection{The transfer fidelity under disorder}

\label{disordered_fidelity}

As we mentioned in the introduction, for a set of manufactured quantum
spin chains, the best-case, worst-case, or average fidelities may
be relevant, depending on the situation at hand. When dealing with
disordered chains the particular realization of the disorder present
in the chain is unknown, unless a complete tomography of the Hamiltonian
can be carried out. Since the situation we imagine is a future ``integrated
quantum circuit'' with a large number of quantum gates and a correspondingly
large number of spin chains connecting gates and registers, the complete
tomography of the multi-chain Hamiltonian is extremely cumbersome.
In that case the average fidelity will be most relevant in order to
compare different protocols and obtain general results. Therefore,
we consider the average of the fidelity (\ref{eq:Averaged-Fidelity}),
evaluated at time $\tau$, over $N_{av}$ realizations of the disorder,
\begin{equation}
\overline{F}(\tau)=\left\langle F(\tau)\right\rangle _{N_{av}}.
\end{equation}
All following numerical simulations employ $N_{av}=10^{3}$ realizations
for the disorder.

\begin{figure}
\centering{}\includegraphics[scale=0.24]{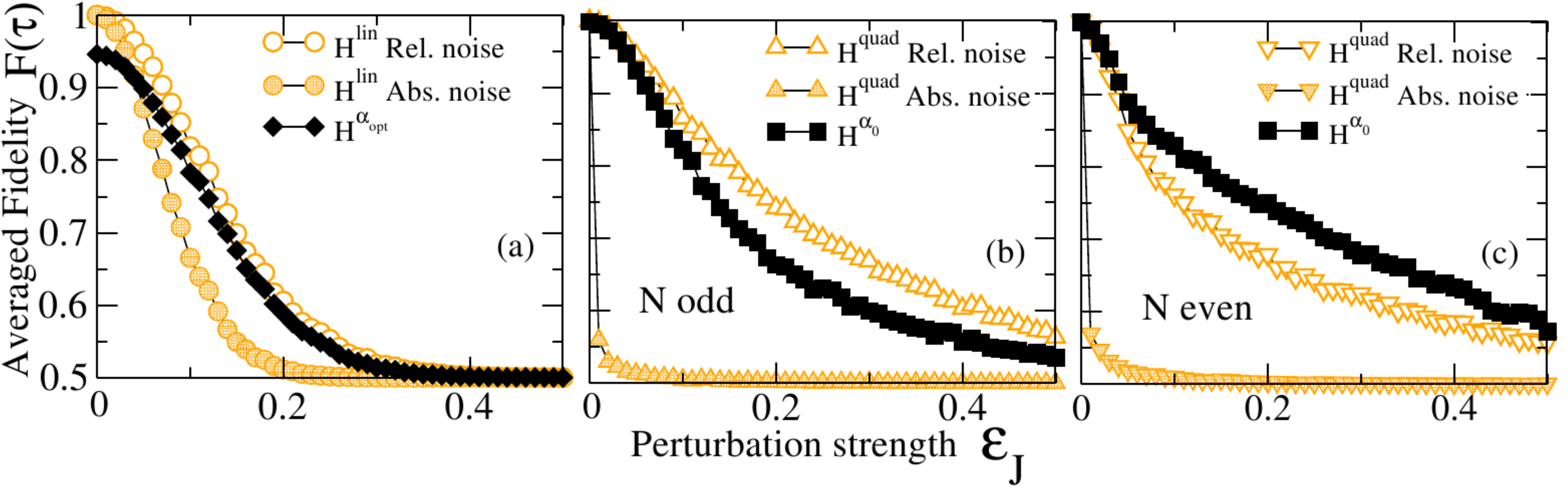} \caption{\label{decaying-fidelity}(Color online) Averaged fidelity at time
$\tau$ as a function of the perturbation strength $\varepsilon_{J}$
for OST and PST channels for a given chain length $N$. Relative and
absolute static disorder are considered. (a) $\overline{F}{}^{lin}$
with relative disorder (open circles) and absolute disorder (filled
orange circles) and $\overline{F}{}^{\alpha_{opt}}$ (black diamonds)
for both kinds of disorder when $N=200$. (b) $\overline{F}{}^{quad}$
with relative disorder (open triangles) and absolute disorder (filled
orange triangles) and $\overline{F}{}^{\alpha_{0}}$ (black squares,
$\alpha_{0}=0.01$) for both kinds of disorder when $N=200$. (c)
Same as panel (b) for $N=201$. The fidelity was averaged over $N_{av}=10^{3}$
realizations of the disorder in each case.}
\end{figure}

Figure \ref{decaying-fidelity} shows the typical behavior of the
averaged fidelity, $\overline{F}(\tau)$, for the different kinds
of channels and disorders here considered, as a function of the disorder
strength. For a fixed length $N$, the averaged fidelity $\overline{F}(\tau)$
is a decreasing function of the disorder strength. The panels are
grouped according to the similarities observed in the spectral properties
of the different systems as discussed above. Obviously, these similarities
are reflected in the performance of the transfer fidelity when relative
noise is considered. For absolute noise, the disorder is more detrimental
for PST systems since $\frac{J_{max}}{J_{min}}$ can be a large number.
A detailed analysis of Fig. \ref{decaying-fidelity} and the comparison
of robustness in these systems follows below.

\begin{figure}
\centering{}\includegraphics[scale=0.6]{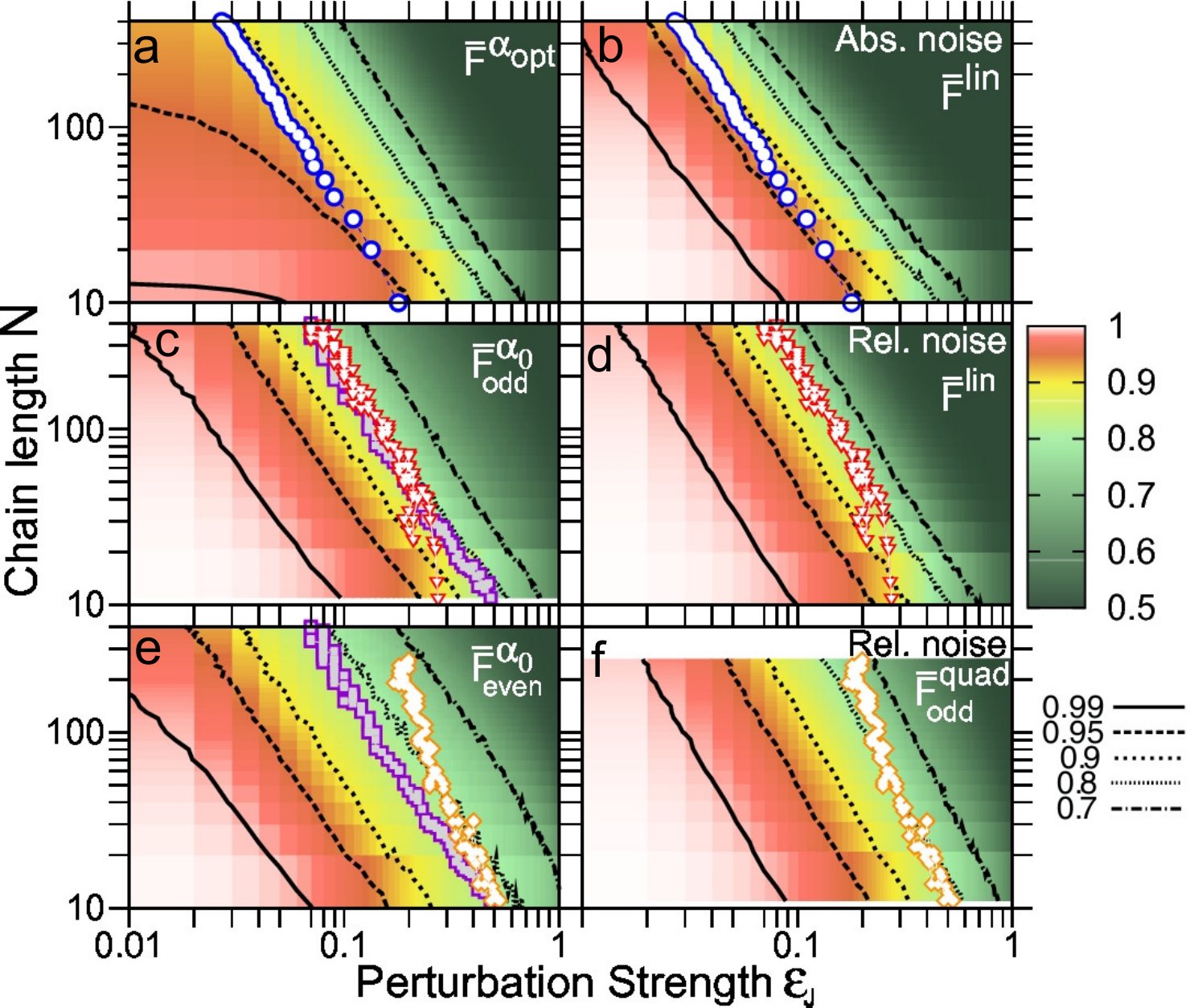} \caption{\label{fidelity-surface}(Color online) Averaged fidelity $\overline{F}{}^{\alpha_{opt}}$,$\overline{F}{}_{even}^{\alpha_{0}}$,$\overline{F}{}_{odd}^{\alpha_{0}}$,
and $\overline{F}{}^{lin}$ when both kinds of disorder are considered
and $\overline{F}{}_{odd}^{quad}$ when only \emph{relative disorder}
is considered.\textbf{ }The average is calculated over $N{}_{av}\!=\!10^{3}$
realizations in each case. The black contour lines belong to $F\!=\!0.99,\,0.95,\,0.9,\,0.8,\,0.7$,
respectively. The colored symbols show the crossovers between the
different systems as explained in the text.}
\end{figure}

Figure \ref{fidelity-surface} shows the averaged fidelity as a function
of the disorder strength $\varepsilon_{J}$ and of the chain length
$N$ for all the channels considered.

The contour lines are defined by $\overline{F}(\tau)=const.$ values.
With the exception of the region where $\overline{F}{}^{\alpha_{opt}}>0.9$,
we observe a very general behavior of the averaged fidelity for the
region of interest: the curves $\overline{F}=const.$ are straight
lines in the logarithmic $(\varepsilon_{J},N)$ plane, \emph{i.e.}
the contour lines of the surface $\overline{F}(\tau)\left[\varepsilon_{J},N\right]$
are curves given by $N\varepsilon_{j}^{\beta}=const.$. We did not
attempt to derive general scaling laws for the systems studied here.
Instead we provide a simple phenomenological description of the numerical
results in a certain range of perturbation strengths and chain lengths.
That range is determined by fidelity values which might be relevant
for state transfer, thus allowing to quantitatively compare the performances
of different channels. Thus, by fitting the averaged fidelities in
this region of interest, we found for all the channels involved in
Fig. \ref{decaying-fidelity}, the scaling function 
\begin{equation}
\overline{F}(N,\varepsilon_{J})=\overline{F}(N\varepsilon_{J}^{\beta})=\frac{1}{2}\left[1+e^{-cN\varepsilon_{J}^{\beta}}\right],\label{scaling}
\end{equation}
where $c$ is a positive constant. Table \ref{tab:decay-exponents}
gives the values of the exponent $\beta$ and the prefactor $c$ for
the different channels. We note that for strong perturbations the
spoiled transfer can be associated to dynamical localization effects
(e.g. Anderson localization) \cite{Stolz_DynamicalLocalization_2012,BEO09,BO07}.
These localization effects are related to zero-velocity Lieb Robinson
bounds, which in XX chain systems lead to the bound $|f_{N}(t)|\leq Ce^{-\eta(\varepsilon_{J})d(N)}$,
where $\eta(\varepsilon_{J})$ is associated with the disorder strength
and $d(N)$ with the site-distance between boundary spins \cite{Stolz_DynamicalLocalization_2012}.
That yields a functional form (similar to (\ref{scaling})) for a
bound of the averaged fidelity $\overline{F}(t)\leq\frac{1}{2}+\frac{Ce^{-\eta(\varepsilon_{J})d(N)}}{3}+\frac{C^{2}e^{-2\eta(\varepsilon_{J})d(N)}}{6}$.

The scaling function (\ref{scaling}) for $\overline{F}{}^{lin}$
with values $\beta=2$ and $c=\frac{1}{5}$ has been reported previously
in Ref. \citep{chiara_perfect_2005}, derived from a perturbative
analysis. The deviation of the numerical values $F^{\alpha_{opt}}>0.9$
from the simple scaling function (\ref{scaling}) can be understood
remembering that the \emph{optimal} channels do not ever achieve $\overline{F}=1$.
Note that $\overline{F}{}_{even}^{\alpha_{0}}$ starts to deviate
faster than $\overline{F}{}_{odd}^{\alpha_{0}}$ from the straight
lines when the disorder strength is reduced for large $N$. That can
be attributed to the fast oscillations described in Sec. \ref{sub:The-transfer-time}
due to the off-resonance transmission. The oscillations induce fluctuations
over realizations that are around twice the oscillation's amplitude,
$2A_{\alpha_{0}}$, which increases with $N$.

\begin{table}
\centering{}%
\begin{tabular}{c|c|c|c|c|c|c|c|}
\cline{2-8} 
 & \multicolumn{3}{c|}{{\small{$\alpha-OST$channels}}} & \multicolumn{4}{c|}{{\small{$PST$channels}}}\tabularnewline
\cline{2-8} 
 & {\small{$F^{\alpha_{opt}}$}} & {\small{$F_{odd}^{\alpha_{0}}$}} & {\small{$F_{even}^{\alpha_{0}}$}} & \multicolumn{2}{c|}{{\small{$F^{lin}$}}} & {\small{$F_{odd}^{quad}$}} & {\small{$F_{even}^{quad}$}}\tabularnewline
\cline{2-8} 
 & \multicolumn{3}{c|}{{\small{Abs. and Rel. noise}}} & {\small{Rel. noise}} & {\small{Abs. noise}} & \multicolumn{2}{c|}{{\small{Rel. noise}}}\tabularnewline
\hline 
\multicolumn{1}{|c|}{{\small{$c$}}} & {\small{0.21}} & {\small{0.17}} & {\small{0.29}} & {\small{0.21}} & {\small{0.20}} & {\small{0.26}} & {\small{0.51}}\tabularnewline
\hline 
\multicolumn{1}{|c|}{{\small{$\beta$}}} & {\small{1.907}} & {\small{1.887}} & {\small{1.81}} & {\small{2.00}} & {\small{1.634}} & {\small{2.14}} & {\small{1.89}}\tabularnewline
\hline 
\end{tabular}\caption{\label{tab:decay-exponents} Values of the constant $c$ and the exponent
$\beta$ for the scaling law $\overline{F}(N\varepsilon_{J}^{\beta})=\frac{1}{2}\left[1+e^{-cN\varepsilon_{J}^{\beta}}\right].$
These values come from the fit parameters of the contour lines displayed
in Fig. \ref{fidelity-surface}, which then are averaged to obtain
a representative value for each of the different systems. Only the
contour lines that can be well fitted by a straight line are considered.
Thus, the scaling law for $F^{\alpha_{opt}}$ is defined by considering
the contour lines with $F^{\alpha_{opt}}\leq0.8$.}
\end{table}

The data analyzed so far in this section suggest that there is no
single simple answer to the question which spin chain is most adequate
to achieve quantum state transfer for a given disorder model. In the
following we try to answer this question in some more detail.

\subsubsection*{Optimal coupling regime\emph{ $\alpha_{opt}$}-OST vs. linear-PST
channels}

Considering \textit{relative} disorder, the main difference between
the systems occurs in the region $\varepsilon_{J}\sim0$, as can be
well observed in Fig. \ref{decaying-fidelity}a, because the case
$\alpha_{opt}$-OST does not produce PST in the limit of zero disorder,
i.e., $\overline{F}{}^{lin}\gtrsim\overline{F}{}^{\alpha_{opt}}$.
For stronger disorder, both fidelities are similar and thus engineering
might not be necessary in that regime. The fidelity's functional dependence
on $\varepsilon_{J}$ and $N$ for the \emph{linear}-PST channel is
$\overline{F}{}^{lin}(N\varepsilon_{J}^{2})\approx\frac{1}{2}\left[1+e^{-\frac{1}{5}N\varepsilon_{J}^{2}}\right]$
\citep{chiara_perfect_2005} as shown in Fig. \ref{fidelity-surface}d.
In contrast, the $\alpha_{opt}$-OST channel shows two regimes depending
on $\varepsilon_{J}$. Within the range of $\varepsilon_{J}$ and
$N$ covered in Fig. \ref{fidelity-surface}, the boundary between
the two regimes is given by $\varepsilon_{J}^{0}\approx(\frac{1.65}{N})^{0.66}$
and a fidelity value of $\overline{F}{}_{\varepsilon_{J}^{0}}^{\alpha_{opt}}=0.91$.
For $\varepsilon_{J}>\varepsilon_{J}^{0}$ the average fidelity of
the $\alpha_{opt}$-OST system scales with $N\varepsilon_{J}^{\beta}$
but $\beta$ varies from $1.61$ to $1.91$. For smaller disorder,
$\varepsilon_{J}<\varepsilon_{J}^{0}$, the behavior is different
as it should, since the $\alpha_{opt}$ system never reaches perfect
fidelity at zero disorder. The fidelity difference between the two
systems at $\varepsilon_{J}^{0}$ is quite small: $\Delta\overline{F}=\overline{F}{}_{\varepsilon_{J}^{0}}^{lin}-\overline{F}{}_{\varepsilon_{J}^{0}}^{\alpha_{opt}}\approx\frac{1}{100}(3\log N-2)$;
in numbers that means $0.004\leq\Delta\overline{F}\leq0.058$.

\begin{figure}[H]
\centering{}\includegraphics[scale=0.19]{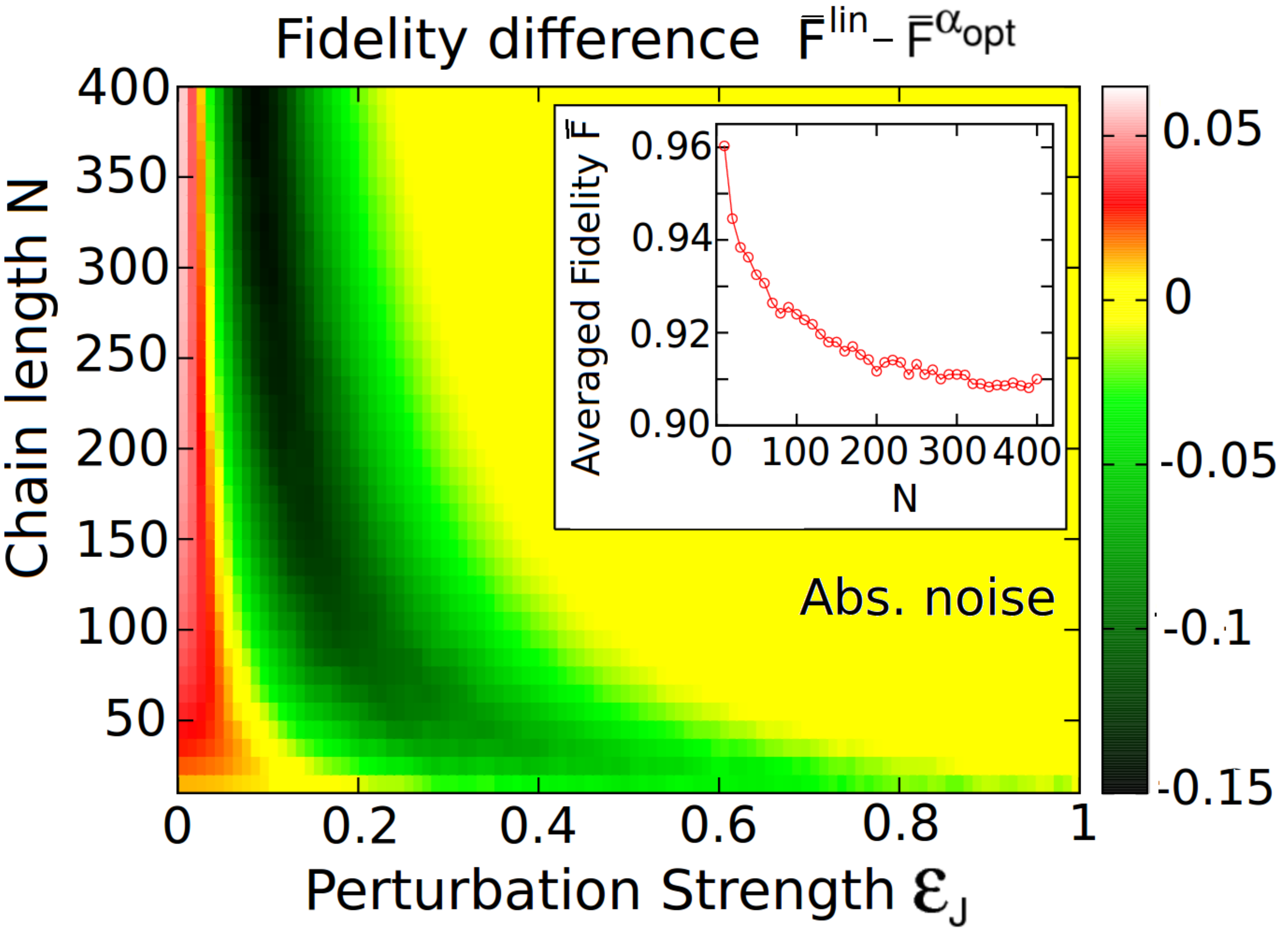} \caption{\label{fig:Difference-fidelity-Flin-FalphOpt}(Color online) Averaged
fidelity difference $\Delta\overline{F}\!=\!\overline{F}{}^{lin}\!-\!\overline{F}{}^{\alpha_{opt}}$
at time $\tau$ as a function of the perturbation strength $\varepsilon_{J}$
and the chain length $N$, averaged over $N_{av}\!=\!10^{3}$ realizations
and with \emph{absolute disorder}. The points where $\Delta\overline{F}\!=\!0$
in the $(N,\varepsilon_{J})$ plane are shown as circles in Figs.
\ref{fidelity-surface}a and \ref{fidelity-surface}b. The inset shows
the value of the fidelity at the crossing point $\Delta\overline{F}\!=\!0$,
as a function of $N$.}
\end{figure}

Considering now \textit{absolute} static disorder, in Fig. \ref{decaying-fidelity}a
we can see that the \textit{linear}-PST system performs better than
$\alpha_{opt}$-OST only for weak perturbations $\varepsilon_{J}$.
For stronger perturbations the $\alpha_{opt}$-OST system overcomes
the\textit{ linear}-PST performance. Studying this behavior as a function
of $N$ as shown in Fig. \ref{fidelity-surface}a and b, we can see
that the crossing point when $\overline{F}{}^{lin}-\overline{F}{}^{\alpha_{opt}}=0$
is determined by $N\varepsilon_{J}^{1.91}\approx0.43$ shown as empty
blue circles in Fig. \ref{fidelity-surface}. For the \emph{linear}-PST
channel with absolute disorder the contour lines of the fidelity are
given by $N\varepsilon^{1.63}$ instead of $N\varepsilon^{2}$ in
the case of relative disorder, while for the minimally engineered
$\alpha_{opt}$-OST channel the scaling behavior remains the same,
as argued earlier.

Figure \ref{fig:Difference-fidelity-Flin-FalphOpt} shows the difference
$\overline{F}{}^{lin}-\overline{F}{}^{\alpha_{opt}}$ and the value
of the fidelity at the crossing point is shown in the inset. It varies
from $0.96$ for $N=10$ to $0.91$, for $N=400$. This demonstrates
that when $N$ increases, the region where the \emph{linear}-PST channel
performs better is reduced and if the perturbation is not small enough,
the minimally engineered $\alpha_{opt}$-OST channel performs better.
As expected, the \emph{linear}-PST channel when suffering \emph{absolute
disorder} is strongly affected as compared to \emph{relative disorder}
where the commensurability of the energy levels is less strongly disturbed
by the disorder. With all of this analysis, we can identify regions
in terms of physical quantities and different disorder models where
the $\alpha_{opt}$-OST performs better than the \textit{linear}-PST
transfer or vice versa. While the $\alpha_{opt}$-OST system is always
faster in terms of transfer time, the quality of the transfer is sometimes
lower. However, this difference of fidelity is more appreciable for
small perturbations and short chains. In the other cases, the $\alpha_{opt}$-OST
system can be more robust.

\subsubsection*{Weak coupling regime \emph{$\alpha_{0}$}-OST vs. quadratic-PST channels}

\begin{figure}
\centering{}\includegraphics[scale=0.19]{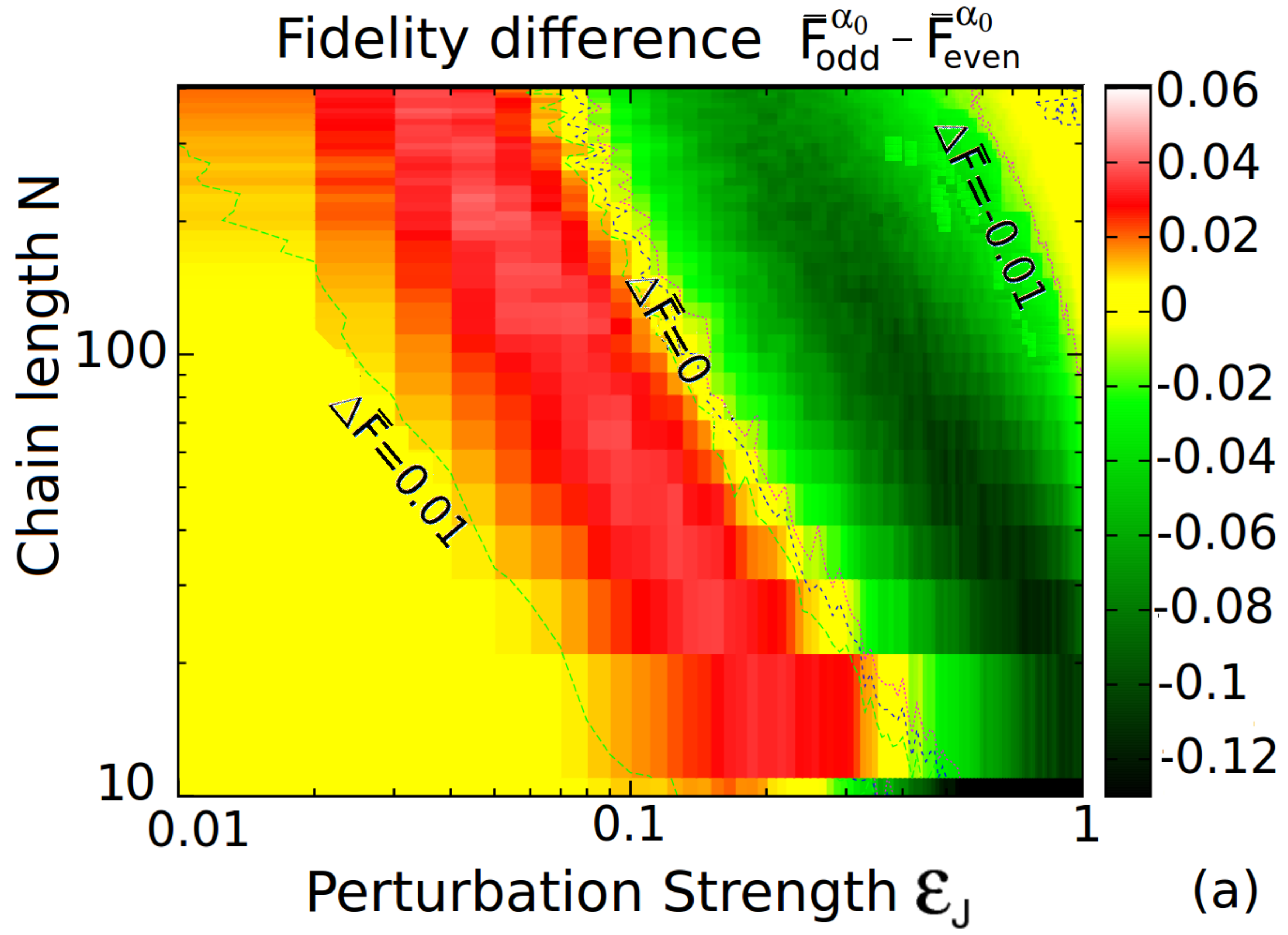}\includegraphics[scale=0.19]{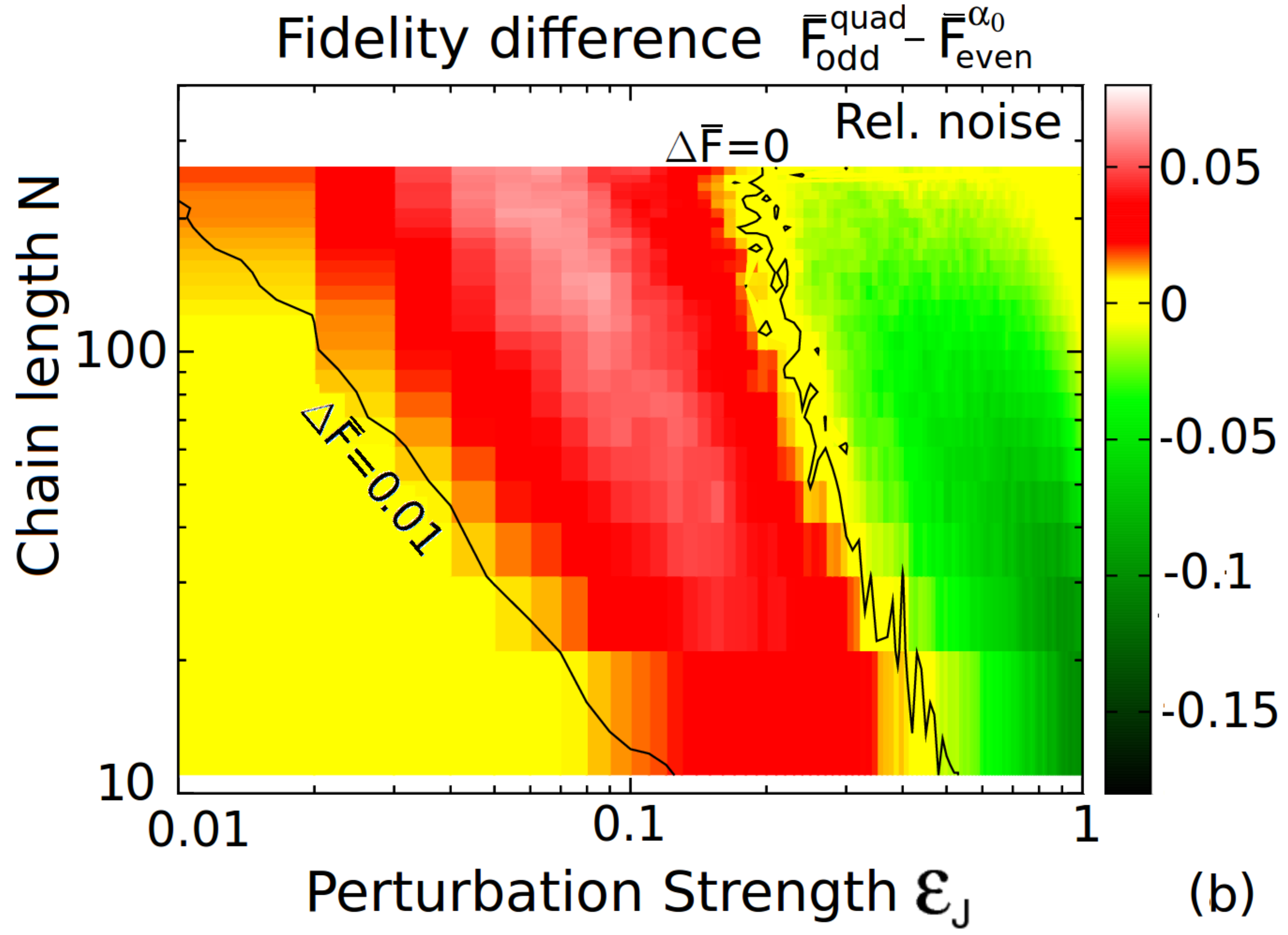}
\caption{\label{fig:Difference-fidelity-Falphaodd-Falph0even-Fquadodd}(Color
online) Averaged fidelity differences $\Delta\overline{F}$ at time
$\tau$ as a function of the perturbation strength $\varepsilon_{J}$
and the chain length $N$, averaged over $N_{av}=10^{3}$ realizations.
\textbf{(a)} $\overline{F}{}_{odd}^{\alpha_{0}}-\overline{F}{}_{even}^{\alpha_{0}}$.
The violet squares in Fig. \ref{fidelity-surface}c and \ref{fidelity-surface}d
show $\Delta\overline{F}(N,\varepsilon_{J})=0$.\textbf{ (b)} $\overline{F}{}_{odd}^{quad}-\overline{F}{}_{even}^{\alpha_{0}}$
with \emph{relative disorder}. The orange diamonds in Fig. \ref{fidelity-surface}e
and \ref{fidelity-surface}f display $\Delta\overline{F}(N,\varepsilon_{J})=0$.}
\end{figure}
\begin{figure}
\centering{}\includegraphics[scale=0.35]{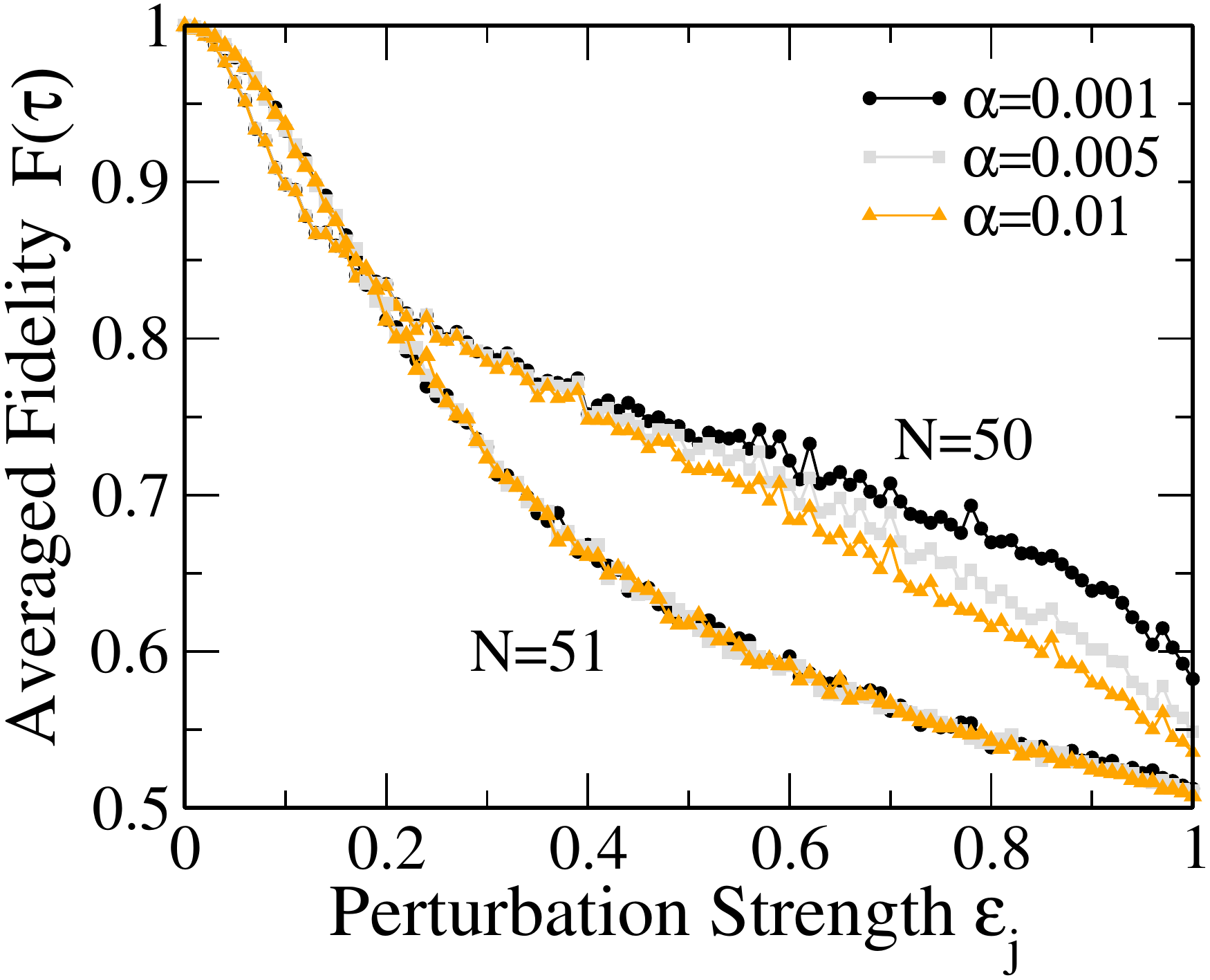} \caption{\label{FalphaO_Neven-odd}(Color online) Averaged transfer fidelity
$F^{\alpha_{0}}$ as a function of the perturbation strength $\varepsilon_{J}$
and the coupling strength $\alpha_{0}$ for relative disorder. The
dependence on $\alpha_{0}$ is only appreciable for even number of
spins $N$ in the chain.}
\end{figure}
Considering \emph{relative disorder}, the fidelity of the boundary-controlled
$\alpha_{0}$-OST system $\overline{F}{}^{\alpha_{0}}$ is similar
or higher (lower) than that of the \emph{quadratic}-PST channel $\overline{F}{}^{quad}$
when $N$ is even (odd), \textit{ie.} $\overline{F}{}_{even}^{\alpha_{0}}\gtrapprox\overline{F}{}_{even}^{quad}$
and $\overline{F}{}_{odd}^{\alpha_{0}}\lessapprox\overline{F}{}_{odd}^{quad}$
(see Figs. \ref{decaying-fidelity}b and \ref{decaying-fidelity}c).
The better performances $\overline{F}{}_{even}^{\alpha_{0}}$ and
$\overline{F}{}_{odd}^{quad}$ are shown in Fig. \ref{fidelity-surface}e
and \ref{fidelity-surface}f respectively. The orange diamond symbols
there indicate where these fidelities are equal, \textit{ie.} $\overline{F}{}_{even}^{\alpha_{0}}=\overline{F}{}_{odd}^{quad}$,
a fit gives $N\varepsilon_{J}^{3.5}\approx0.62$ and the fidelity
value is around $F\approx0.8$, decreasing for larger $N$. To the
left of the symbols $\overline{F}{}_{odd}^{quad}>\overline{F}{}_{even}^{\alpha_{0}}$,
but for small perturbation strength the differences between the two
systems are quite small as visible in Fig. \ref{fig:Difference-fidelity-Falphaodd-Falph0even-Fquadodd}b.

The fidelity contour lines $\overline{F}{}_{even}^{\alpha_{0}}=\mathrm{const}$
do not follow the scaling $N\varepsilon_{J}^{\beta}\sim\mathrm{const}$
for values below $0.8$. In this region, the effect of the perturbation
depends on $\alpha_{0}$, but only for even $N$. That is shown in
Fig. \ref{FalphaO_Neven-odd} which also demonstrates marked differences
between even and odd $N$ in the dependence of $\overline{F}^{\alpha_{0}}(\tau)$
on $\varepsilon_{J}$. Note, however, that the differences are most
conspicuous in the region where the fidelity is much too small anyway
for reliable quantum information processing.

Considering \emph{absolute disorder}, the minimally engineered $H^{\alpha_{0}}$
systems are always the most robust ones for transferring information.
The fidelity of $H^{quad}$ in that case decays very rapidly as a
function of $N$ and $\varepsilon_{J}$ (not shown). This is connected
to the fact that the maximum and minimum couplings, $J_{max}^{quad}$
and $J_{min}^{quad}$, in the chain may differ by orders of magnitude,
with the smallest couplings always close to the ends of the chain
$J_{min}^{quad}=J_{1}^{quad}=J_{N}^{quad}$. We found a relation \textbf{$\frac{J_{max}^{quad}}{J_{min}^{quad}}\sim0.06N^{2}$}
for even $N$\textbf{ }and \textbf{$\sim0.1N^{2}$} for odd $N$\textbf{.}
Consequently, a fluctuation of a given absolute size may completely
spoil the state transfer when it affects one of the small couplings
close to the boundary as observed in Figs. \ref{decaying-fidelity}b
and \ref{decaying-fidelity}c for $N=200$.

Returning to the $\alpha_{0}$-OST channels, in Figs. \ref{fidelity-surface}c
and \ref{fidelity-surface}e we can compare the fidelities $\overline{F}{}_{odd}^{\alpha_{0}}$
and $\overline{F}{}_{even}^{\alpha_{0}}$. The violet squares show
when $\overline{F}{}_{odd}^{\alpha_{0}}\!=\!\overline{F}{}_{even}^{\alpha_{0}}$,
where $\overline{F}{}_{odd}^{\alpha_{0}}\!>\!\overline{F}{}_{even}^{\alpha_{0}}$
to the left of the symbols. This crossing line approximately follows
$N\varepsilon_{J}^{1.98}\!\approx\!0.50$ and corresponds to a constant
fidelity value $F^{\alpha_{0}}\!\gtrapprox\!0.8$. The differences
between these fidelities are displayed in Fig. \ref{fig:Difference-fidelity-Falphaodd-Falph0even-Fquadodd}a.

\subsubsection*{Minimally engineered vs. fully engineered channels}

\begin{figure}
\centering{}\includegraphics[scale=0.2]{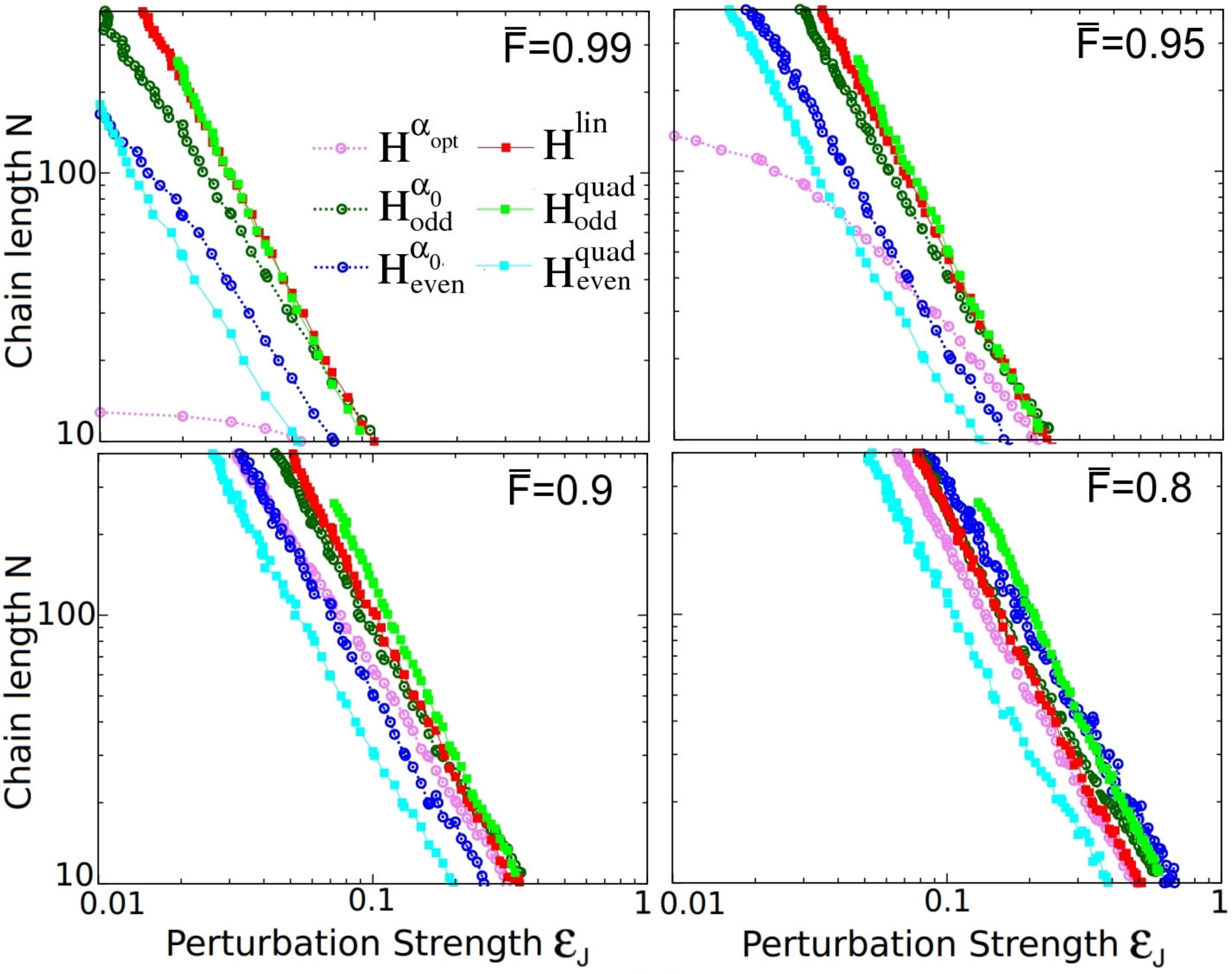} \caption{\label{fig:comparison-Relnoise}(Color online) Contour lines of the
averaged transfer fidelity for fully-engineered perfect state transfer
systems (closed symbols) and boundary-controlled $\alpha$-optimized
state transfer systems (open symbols) for relative noise. They are
shown as a function of the perturbation strength $\varepsilon_{J}$
and the chain length $N$ for the averaged fidelities $\overline{F}=0.99,0.95,0.9,0.8$.}
\end{figure}
Comparing all of the systems for \emph{relative disorder}, for small
perturbation, most of the systems achieve a high fidelity of state
transfer. In general we observe in Fig. \ref{fig:comparison-Relnoise}
that for fidelities $\gtrsim0.8$, where the functional dependence
of the infidelity $1-\overline{F}$ as a function of the perturbation
strength is quadratic for all systems with $\overline{F}\rightarrow1$
for $\varepsilon_{J}\to0$, the most robust system is $H_{odd}^{quad}$
and the least robust one $H_{even}^{quad}$. This means that the contour
lines in the $(N,\varepsilon_{J})$ plane are bounded by the ones
belonging to $H_{odd}^{quad}$ from above and by the $H_{even}^{quad}$
from below. For $F>0.8$ the most robust system is $H_{odd}^{quad}$,
but the $H^{linear}$ system follows with very similar performance,
although the $H_{odd}^{\alpha_{0}}$ system is also comparable since
their fidelities, $\overline{F}{}^{lin}$ and $\overline{F}{}_{odd}^{\alpha_{0}}$,
differ only by 4\% or less as $\varepsilon_{J}$ and $N$ increase.
In Fig. \ref{fidelity-surface}c and \ref{fidelity-surface}d, the
red down-triangle symbols show where $\overline{F}{}^{lin}=\overline{F}{}_{odd}^{\alpha_{0}}\approx0.8$.
When $N\varepsilon_{J}^{1.93}\gtrsim2.58$, $\overline{F}{}_{odd}^{\alpha_{0}}>\overline{F}{}^{lin}$.
The different behaviors and regimes just mentioned can be seen from
Fig. \ref{fig:comparison-Relnoise} where contour lines corresponding
to several values of the averaged fidelities for all the systems are
shown. 

\begin{figure}[H]
\centering{}\includegraphics[scale=0.18]{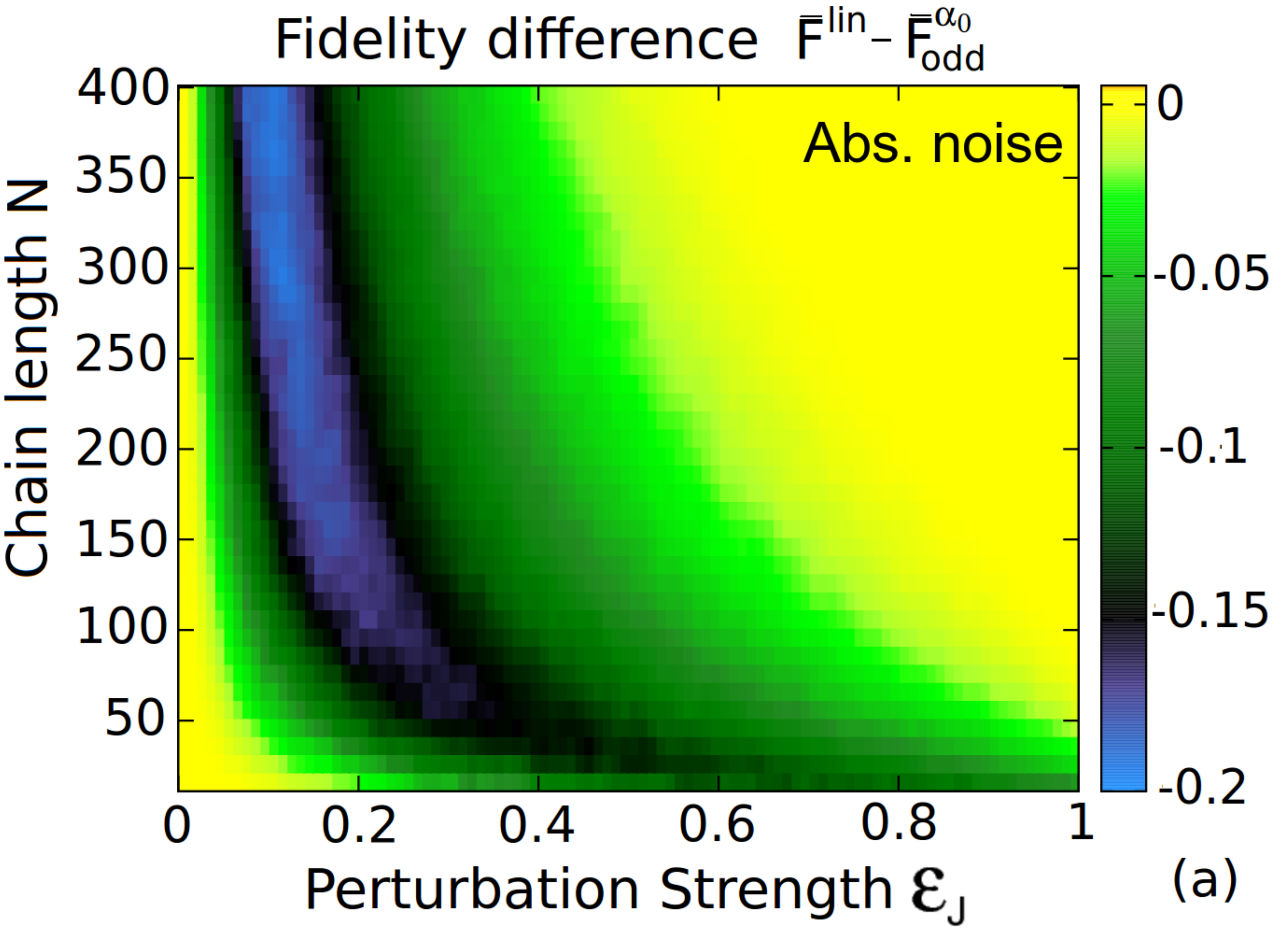}\includegraphics[scale=0.18]{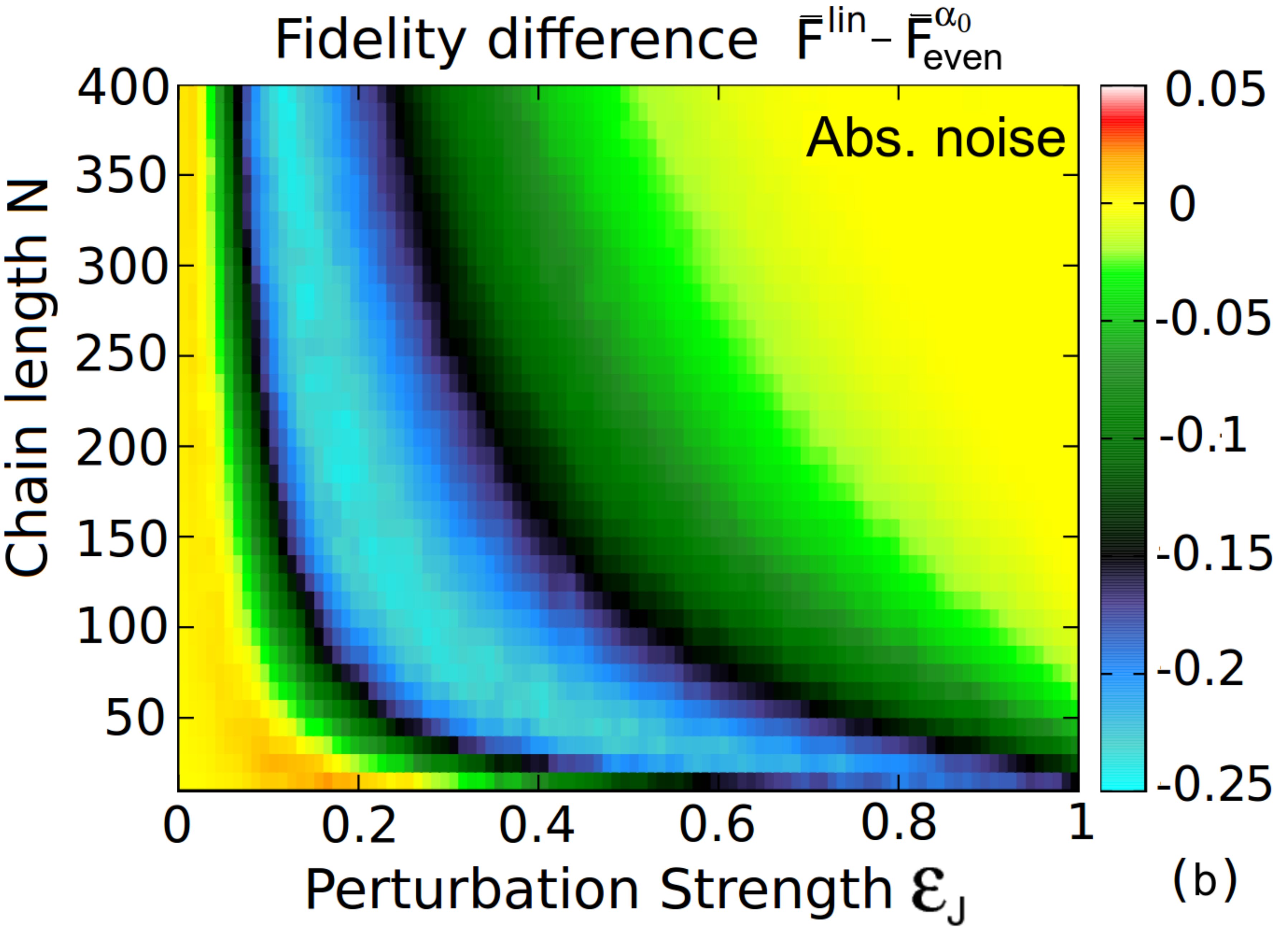}
\caption{\label{fig:Difference-fidelity-Flin-Falph0odd-alphOpt}(Color online)
Averaged fidelity differences $\Delta\overline{F}$ at time $\tau$
as a function of the perturbation strength $\varepsilon_{J}$ and
the chain length $N$, averaged over $N_{av}=10^{3}$ realizations.\textbf{
}\emph{Absolute disorder} is considered for \textbf{(a)} $\overline{F}{}^{lin}-\overline{F}{}_{odd}^{\alpha_{0}}$
and \textbf{(b)} $\overline{F}{}^{lin}-\overline{F}{}_{even}^{\alpha_{0}}$. }
\end{figure}

For \emph{absolute disorder}, the minimally engineered quantum channels
are clearly most robust against perturbations. In this case, $H^{\alpha_{0}}$
is most robust. See the previous section for the comparison of the
fidelities $\overline{F}{}_{odd}^{\alpha_{0}}$ and $\overline{F}{}_{even}^{\alpha_{0}}$.
Only for small perturbation the best $\overline{F}{}^{lin}$ performance
is similar to $\overline{F}{}^{\alpha_{0}}$ as can be seen in Figs.
\ref{fig:Difference-fidelity-Flin-Falph0odd-alphOpt}a and \ref{fig:Difference-fidelity-Flin-Falph0odd-alphOpt}b.

In terms of the transfer time, the fastest transfer is achieved almost
at the quantum speed limit \citep{murphy_communication_2010} by the
$\alpha_{opt}$-OST system, then follows the \emph{linear}-PST system,
and the transfer times of the remaining systems depend on the values
of $N$ and $\alpha$, where always $\tau_{odd}^{\alpha_{0}}\leqslant\tau_{even}^{\alpha_{0}}$
and for $\alpha\lesssim\frac{8\sqrt{N-2}}{N^{3/2}}$, $\tau^{quad}<\tau_{odd}^{\alpha_{0}}$.

\section{Conclusions}

\label{conclusions}

We studied the robustness of nearest-neighbor coupled spin chains
for state transfer against different kinds of static coupling-constant
disorder. We identified perfect state transfer (PST) channels that
are robust against static disorder in the coupling strength. We found
and showed similarities in the spectral properties that are responsible
for the dynamics of the transfer for some minimally engineered channels
on the one hand, those that we call boundary-controlled channels,
and fully engineered PST channels on the other hand. We showed that
these minimally engineered systems perform similarly to fully engineered
systems. We conclude that in many situations, the minimally engineered
channels are similar to or even more efficient and robust in presence
of perturbations than the fully engineered channels. This points out
possibilities to circumvent the obvious difficulties inherent in implementing
experimentally a fully engineered system with nearest-neighbor couplings
which vary over orders of magnitude within the system. Moreover, one
of the minimally engineered systems (the $\alpha_{opt}$-OST channel)
achieves the fastest state transfer.

Additionally, we documented a common decay law for the transfer fidelity,
$\overline{F}(N\varepsilon_{J}^{\beta})=\frac{1}{2}\left[1+e^{-cN\varepsilon_{J}^{\beta}}\right]$,
as a function of the perturbation strength $\varepsilon_{J}$ and
the channel length $N$, where the exponent $\beta$ turns out to
be close to $2$ for all systems in the parameter range studied for
the different systems. This law quantifies the sensitivity and robustness
against perturbations. The parameters of these scaling laws which
we provide for the different proposed quantum channels can serve to
judge which configuration would be optimal for realizing state transfer
in a given situation. While the exponent $\beta=2$ can be derived
from a perturbative treatment \cite{chiara_perfect_2005}, there is
no straightforward explanation for the deviations from that integer
exponent value. On one hand, for the (minimally engineered) OST cases,
the fidelity scaling law may deviate from that simple form due to
the fidelity not being perfect for vanishing disorder strength. On
the other hand, for strong disorder, dynamical localization effects
may influence the scaling laws \cite{Stolz_DynamicalLocalization_2012}.

% \newpage{}

\section{Acknowledgements}

A. Z. and O. O. acknowledge support from SECYT-UNC and CONICET. A.Z.
thanks for support by DAAD. GAA acknowledges the support of the European
Commission under the Marie Curie Intra-European Fellowship for career
Development.

\appendix
%dummy comment inserted by tex2lyx to ensure that this paragraph is not empty
%dummy comment inserted by tex2lyx to ensure that this paragraph is not empty
%dummy comment inserted by tex2lyx to ensure that this paragraph is not empty

\section{Gaussian distribution of $P_{k,1}^{lin}$\label{sec: Appendix B. Gaussian Pk1}}

The normalized eigenstates of a chain of $N+1$ spins 1/2 with XX-type
couplings between sites $l$ and $l+1$ given by 
\begin{equation}
J_{l}=\sqrt{(l+1)(N-l)}
\end{equation}

are Krawtchouk polynomials \cite{albanese_mirror_2004}. The amplitude
of the ground state at site $l(=0,1,...,N)$ is 
\begin{equation}
\phi_{0}(l)=\sqrt{\frac{1}{2^{N}}\left(\begin{array}{c}
N\\
l
\end{array}\right)},
\end{equation}
that is, the square root of the binomial distribution, 
\begin{equation}
b(l,N,p)=\left(\begin{array}{c}
N\\
l
\end{array}\right)p^{k}(1-p)^{N-k}\label{eq:binomial}
\end{equation}
for $p=\frac{1}{2}$, with expectation value $\langle l\rangle=Np=\frac{1}{2}$
and variance $\sigma^{2}=Np(1-p)=\frac{N}{4}$ . For large values
of N the binomial distribution approaches the Gaussian distribution
(with the same expectation value and variance) in the range where
the probability is non-negligible; that is, 
\begin{equation}
|\phi_{0}(l)|^{2}\approx\sqrt{\frac{2}{N\pi}}\exp{\left(-\frac{2}{N}\left(l-\frac{N}{2}\right)^{2}\right)}.\label{eq:gaussianapprox}
\end{equation}

\begin{figure}[floatfix]
\begin{centering}
\includegraphics[scale=0.28]{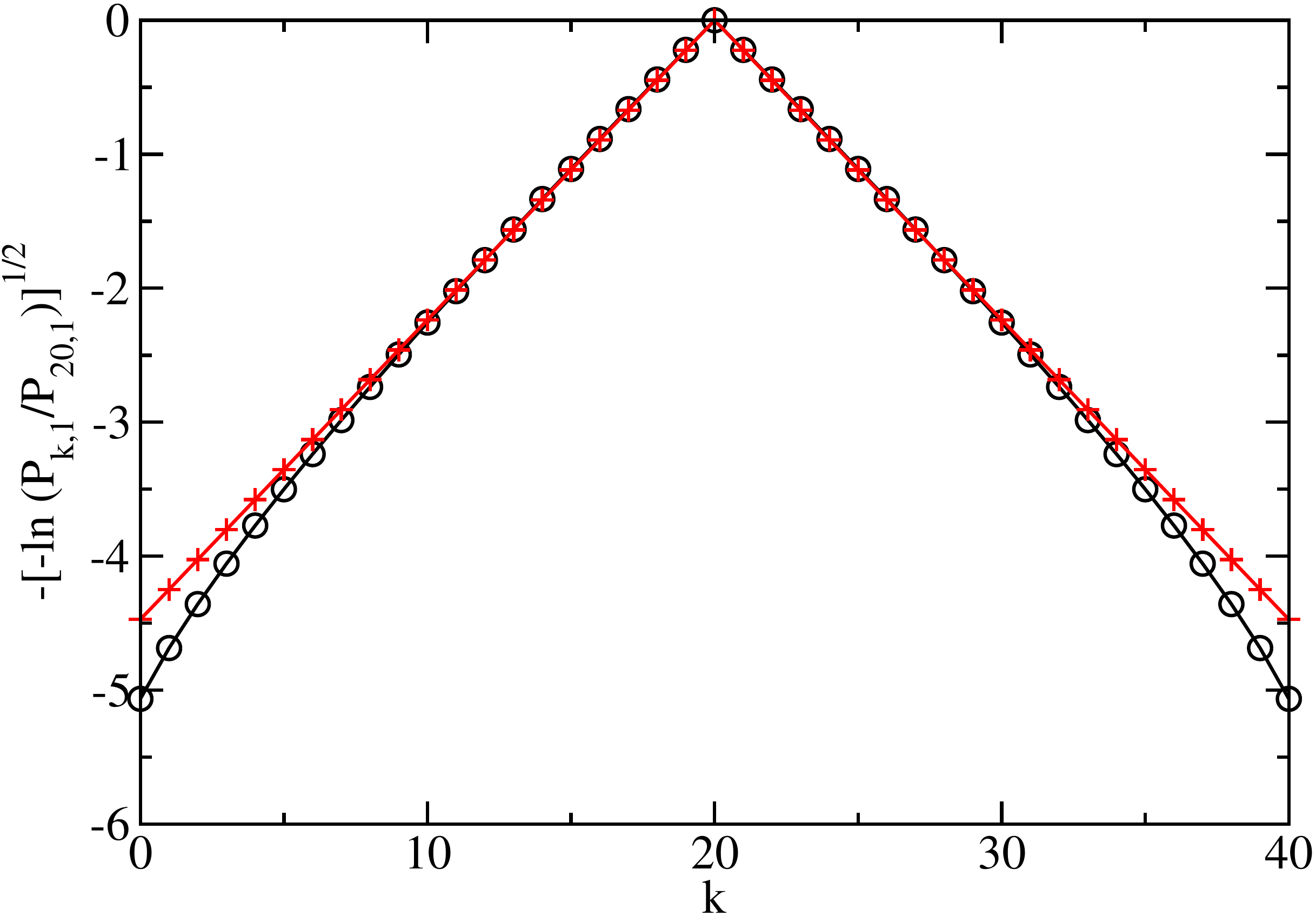} 
\par\end{centering}

\centering{}\caption{\label{fig:pk1approx}(Color online) The probabilities $P_{k,1}$
obtained using the exact expression (\ref{eq:binomial}) (black circles)
and the Gaussian approximation (\ref{eq:gaussianapprox}) (red plus
signs), for a 41-site chain with sites labeled $k=0,1,...,40$. Plotted
is minus the square root of $-\ln(P_{k,1}/P_{20,1})$, so that the
Gaussian is represented by two straight lines.}
\end{figure}

Figure~\ref{fig:pk1approx} shows the probabilities $P_{k,1}$ calculated
using the exact amplitudes, Eq.~\ref{eq:binomial}, and the Gaussian
approximation given by Eq.~\ref{eq:gaussianapprox}, for a spin chain
with $N=40$ (that is, with 41 spins). Obviously the agreement between
the exact and approximate probabilities is excellent particularly
at, and near, the center of the band of eigenvalues.

\addcontentsline{toc}{section}{\refname}\bibliographystyle{ChemEurJ}
s \bibliographystyle{ChemEurJ}
\bibliography{OST,OST_JS}

\end{document}